\documentclass[aps,twocolumn,groupedaddress]{revtex4}

\usepackage{epsfig}
\usepackage{amsmath}
\usepackage{braket}
\usepackage{xspace}
\usepackage{graphicx}
\usepackage{lineno}

\begin{document}
\draft
%\twocolumn[\hsize\textwidth\columnwidth\hsize\csname@twocolumnfalse\endcsname

\title{Classifying coherent peaks in nanoelectronic devices by the presence or absence of spin exchange}
\author{Jongbae Hong}
\affiliation{School of Physics and Astronomy, Seoul National University, 1 Gwanak-ro, Gwanak-gu, Seoul 08826, Korea}

\date{\today}

\begin{abstract}
Coherent peaks appearing in the differential conductance of quantum-dot single-electron transistors (QDSETs) and quantum point contact (QPC) devices are classified into two categories according to the scaling function onto which the temperature-scaled differential-conductance maxima collapse and the underlying spin dynamics. The zero-bias peaks (ZBPs) observed in QPCs and in the triplet state of the even-particle sector of QDSETs belong to the same category, whereas the ZBP in the odd-particle sector of a QDSET belongs to a different category together with all finite-bias coherent peaks observed in QPCs and in the even-particle sector of QDSETs. 
The spin dynamics of the former category involve spin exchange, a hallmark of Kondo dynamics, whereas those of the latter category involve only cotunneling of an up--down spin pair. 
Furthermore, for the former type of ZBP, the scaling temperature coincides with one-half of the full width at half maximum (FWHM), which corresponds to the Kondo temperature. In contrast, for the latter type, the scaling temperature does not coincide with the (1/2)FWHM-derived energy scale.
To support these findings, the gate-voltage-dependent differential-conductance line shapes measured in the odd-particle sector of a QDSET are theoretically reproduced. 
The results demonstrate that the observed ZBP is a merging of two coherent side peaks generated solely by the cotunneling of up--down spin pairs.
\end{abstract}

%\pacs{72.15.Qm, 73.63.Rt, 73.23.-b, 75.76.+j}

\maketitle \narrowtext 

\section{Introduction}

\begin{figure}[b] 
\centering
\includegraphics[width=3.2 in]{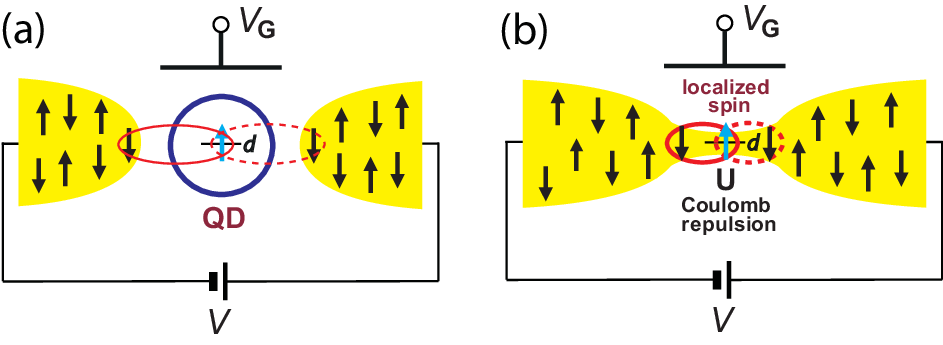}
\caption{Schematic diagrams of two-terminal nanodevices based on a quantum dot (a) and a QPC (b), respectively, forming a left--right entangled state. 
The thick ellipse in (b) indicates a stronger singlet-coupling strength than the thin ellipse in (a). 
$U$ denotes the on-site Coulomb repulsion at the mediating spin occupying level $d$, and $V_{\rm G}$ denotes the side-gate voltage.
}
\label{fig1}
\end{figure}

Quantum technology has become one of the most active research directions in modern advanced industries. 
Advances in nanofabrication have enabled the realization of confined nanostructures such as quantum dot single-electron transistors (QDSETs)~\cite{Goldhaber,Cronenwett98} and quantum point contact (QPC) devices~\cite{Kristensen,Cronenwett}.
The basic structure of the nanodevice considered in this study consists of two metallic reservoirs and a localized spin between them, as illustrated in figure~\ref{fig1}.
The localized spin may reside in an artificially created atom (figure~\ref{fig1}(a)) or emerge spontaneously in a narrow constriction (figure~\ref{fig1}(b))~\cite{Hirose,Rejec,Yoon}.

%%%%%%%%%%%%%%%%%%%%%%%%%%%%%%%%%%%%%%%%%%%%%%%%%%%%%%%%%%%%%%%%%%%%%%%%%%%%%%%%%%%%%%%%
%%%%%%%%%%%%%%%%%%%%%%%%%%%%%%%%%%%%%%%%%%%%%%%%%%%%%%%%%%%%%%%%%%%%%%%%%%%%%%%%%%%%%%%%

At sufficiently low temperatures and bias voltages, electrons in a QDSET undergo cotunneling processes~\cite{Averin}, which can induce spin-flip transitions of the single spin occupying the highest quantum-dot level.
Cronenwett, Oosterkamp, and Kouwenhoven~\cite{Cronenwett98} explain that the successive spin-flip processes through cotunneling are similar to the behavior of a spin-singlet state that generates the Kondo effect by screening the local spin. 
Since the cotunneling in nanoelectronic devices shown in figure~\ref{fig1} is coherent, we refer to the coherent cotunneling process involving an up--down spin pair as {\it singlet cotunneling} in this work.
Such singlet cotunneling processes produce a Kondo-like coherent peak in differential conductance. 
However, this type of peak should be distinguished from coherent peaks that arise from spin-exchange dynamics.
One objective of this work is to establish that distinction using scaling analysis and by identifying the underlying spin dynamics responsible for each peak.

Important problems associated with the scaling analysis of correlated quantum dot systems are finding scaling functions for a large localized spin, such as spin value 1 or 3/2~\cite{Parks,Roch00} and understanding the scaling behavior near the quantum critical point present in multi-channel quantum dot devices~\cite{Iftikhar}.
Recently, an interesting universal magnetic-field scaling in nanoelectronic systems has been reported~\cite{Hsu}. 
Despite the existence of such advanced topics regarding the scaling study for nanodevices, this study focuses on the fundamental phenomenon, in which the scaling function for a coherent peak depends on the spin dynamics used to form that coherent peak.

In quantum devices using quantum dots or QPCs, spin dynamics are classified into two types depending on the presence or absence of spin exchange.
As shown in figure~\ref{fig1}, the linear combination of the two singlets on the left and right forms a left--right entangled state involving all the coherent spins of the system.
Singlet cotunneling is essential in these nanodevices because it is the only way to mediate a coherent current flow between the two reservoirs.

\begin{figure}[t] 
\centering
\includegraphics[width=3.2 in]{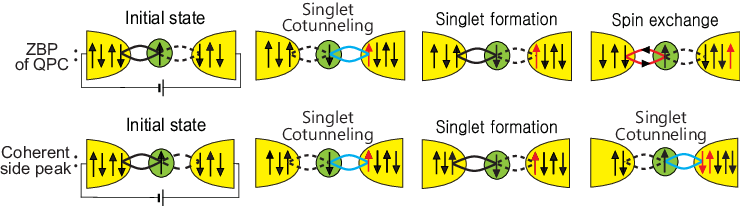}
\caption{Spin dynamics of unidirectional entangled-state tunneling in a QPC device~\cite{iop-qpc}. Dynamics involving spin exchange (top) generate the ZBP, whereas dynamics involving only singlet cotunneling (bottom) generate coherent side peaks.
 }
\label{fig2}
\end{figure}

Unlike the QDSET, the QPC device shown in figure~\ref{fig1}(b) has no potential barrier forming a quantum dot and has a short singlet length, resulting in a much stronger singlet coupling strength than that of the QDSET in figure~\ref{fig1}(a). 
Consequently, spin exchange can occur in QPC devices~\cite{iop-qpc}, as shown in figure~\ref{fig2}, whereas coherent transport is governed solely by cotunneling in QDSETs~\cite{Cronenwett98}.

The primary observable that reflects the underlying dynamics of the system is the differential conductance, $dI/dV$, measured as a function of source--drain bias voltage $V$.
QDSETs made of GaAs/AlGaAs heterostructures~\cite{Goldhaber}, carbon nanotubes (CNTs)~\cite{Nygard}, Aharonov--Bohm rings~\cite{Wiel}, gold particles~\cite{Heersche}, and bilayer graphene with spin--orbit coupling~\cite{Kurzmann} commonly exhibit a central coherent peak at zero bias and two broad incoherent Coulomb-blockade peaks in their differential conductance when the gate voltage lies in the odd-particle sector of the Coulomb diamond formed in a source--drain bias versus gate-voltage plot.

In contrast, QPCs exhibit a relatively narrow and weak zero-bias peak (ZBP) with two additional coherent side peaks, together with two broad Coulomb peaks that are usually not seen in experimental data~\cite{Cronenwett,Chen,Sarkozy,Ren,DiCarlo}.
Recent study on bilayer graphene demonstrates the QPC-type differential conductance and addresses the transition between SU(4) and SU(2) Kondo effects~\cite{Josep}, which may offer new insights into correlated transport phenomena in nanoelectronic devices.

Another interesting phenomenon emerges in the even-particle sector of QDSETs, where coherent side peaks observed in QPC devices appear when the quantum dot hosts a spin triplet, in which case the differential conductance shows a prominent ZBP.
Examples include bilayer graphene with spin--orbit coupling~\cite{Kurzmann} and C$_{60}$ molecules~\cite{Roch}. 
Despite their very different physical realizations, these two QDSETs, both with a spin triplet in the quantum dot, exhibit the same differential-conductance structure, suggesting that the two coherent side peaks do not arise from the specific properties of the quantum dot itself.
This claim is supported by the feasibility of a scaling analysis of the corresponding side peaks shown in reference~\cite{Kurzmann}.

We explain in the last part of section~\ref{sec:reproduct} that the two side peaks are generated under steady-state nonequilibrium conditions and disappear in equilibrium.
We will also show that these differential-conductance structures have the same basis as the three coherent peak structures of QPC devices discussed in our previous study~\cite{iop-qpc}, in which the spin dynamics generating the zero-bias anomaly involve spin exchange, whereas the two coherent side peaks are formed solely by singlet cotunneling, as shown in figure~\ref{fig2}.

%%%%%%%%%%%%%%%%%%%%%%%%%%%%%%%%%%%%%%%%%%%%%%%%%%%%%%%%%%%%%%%%%%%%%%%%%%%%%%%%%%%%%%%%
%%%%%%%%%%%%%%%%%%%%%%%%%%%%%%%%%%%%%%%%%%%%%%%%%%%%%%%%%%%%%%%%%%%%%%%%%%%%%%%%%%%%%%%%

The spin dynamics operating in the two nanosystems shown in figure~\ref{fig1} can be described using the two-reservoir Anderson impurity model~\cite{Hewson}. 
In this study, we reveal which spin dynamics generate which coherent peaks observed in QDSETs and QPC devices when coherent transport occurs through unidirectional entangled-state tunneling under bias.

To achieve the above-mentioned aim, we classify all coherent peaks appearing in typical nanodevices fabricated with QDs and QPCs into groups with different spin dynamics and scaling characteristics.
By different spin dynamics, we mean distinguishing coherent peaks formed solely through singlet cotunneling from those involving spin exchange.
By different scaling characteristics, we mean that different scaling functions are used.
To this end, we employ several tools, including scaling analysis, the relationship between ZBP width and Kondo temperature, and theoretical reproduction of differential-conductance line shapes.
The last tool distinguishes the spin dynamics that form each coherent peak.
The main focus of this work is the ZBP appearing in the odd-particle sectors of QDSETs.

The remainder of this paper is organized as follows. 
Section~\ref{sec:scaling} presents the results of scaling analysis for various coherent peaks in QDSET and QPC devices and classifies them according to their respective scaling functions.
In this section, we compare the scaling temperature with the Kondo temperature.
Section~\ref{sec:formula} introduces the Green's function technique in operator space and derives a convenient formula for the differential conductance, which is used in section~\ref{sec:reproduct} to reproduce the experimental differential-conductance line shape theoretically and to identify the spin dynamics responsible for each coherent peak.
We present our discussion in section~\ref{sec:discuss}.
Section~\ref{sec:conclusion} summarizes our conclusions, followed by detailed appendices.

\section{Scaling analysis and the Kondo temperature} \label{sec:scaling} 

Scaling analysis is a powerful tool for identifying Kondo-related phenomena in mesoscopic systems. 
In QDSET and QPC devices, scaling analyses are typically performed using the temperature dependence of the linear conductance at zero bias. 
In the present work, we extend the scope of the scaling analysis to include finite-bias coherent peaks, which have generally been excluded from previous studies,
with the exception of reference~\cite{Kurzmann}, to the best of our knowledge.
The motivation for including finite-bias coherent peaks is based on the explanation of Cronenwett $\textit{et al.}$~\cite{Cronenwett98} that successive 
singlet-cotunneling processes generate the ZBP in the odd-particle sector of QDSETs, together with our recent finding that finite-bias coherent peaks originate from singlet cotunneling~\cite{iop-qpc}. 

Consequently, all coherent peaks observed in the odd- and even-particle sectors of QDSETs, as well as those observed in QPC devices, are included in the present scaling analysis, enabling a unified classification based on their scaling behavior.
A well-known fact is that the scaled data of the ZBPs in the odd-particle sector of QDSETs collapse onto scaling function~\cite{Goldhaber-PRL}: 
\begin{equation}
G_{\rm I}(T)=G_0\{1+(2^{1/s}-1)(T/T_{\rm SC})^2\}^{-s},
\label{SF-1}
\end{equation}
with $s=0.22$ and zero-temperature conductance $G_0$; and those observed in QPC devices collapse onto scaling function~\cite{Cronenwett}:
\begin{equation}
G_{\rm II}(T)=\frac{2e^2}{h}[0.5\{1+(2^{1/s}-1)(T/T_{\rm SC})^2]^{-s}\}+0.5]. \, \, \,
\label{SF-2}
\end{equation}
Here, $T_{\rm SC}$ denotes the scaling temperature. 
Although it is often identified with the Kondo temperature, we defer adopting this interpretation until the end of this section.

%%%%%%%%%%%%%%%%%%%%%%%%%%%%%%%%%%%%%%%%%%%%%%%%%%%%%%%%%%%%%%%%%%%%%%%%%%%%%%%%%%%%%%%%
\begin{figure}[t] 
\centering
\includegraphics[width=3.2 in]{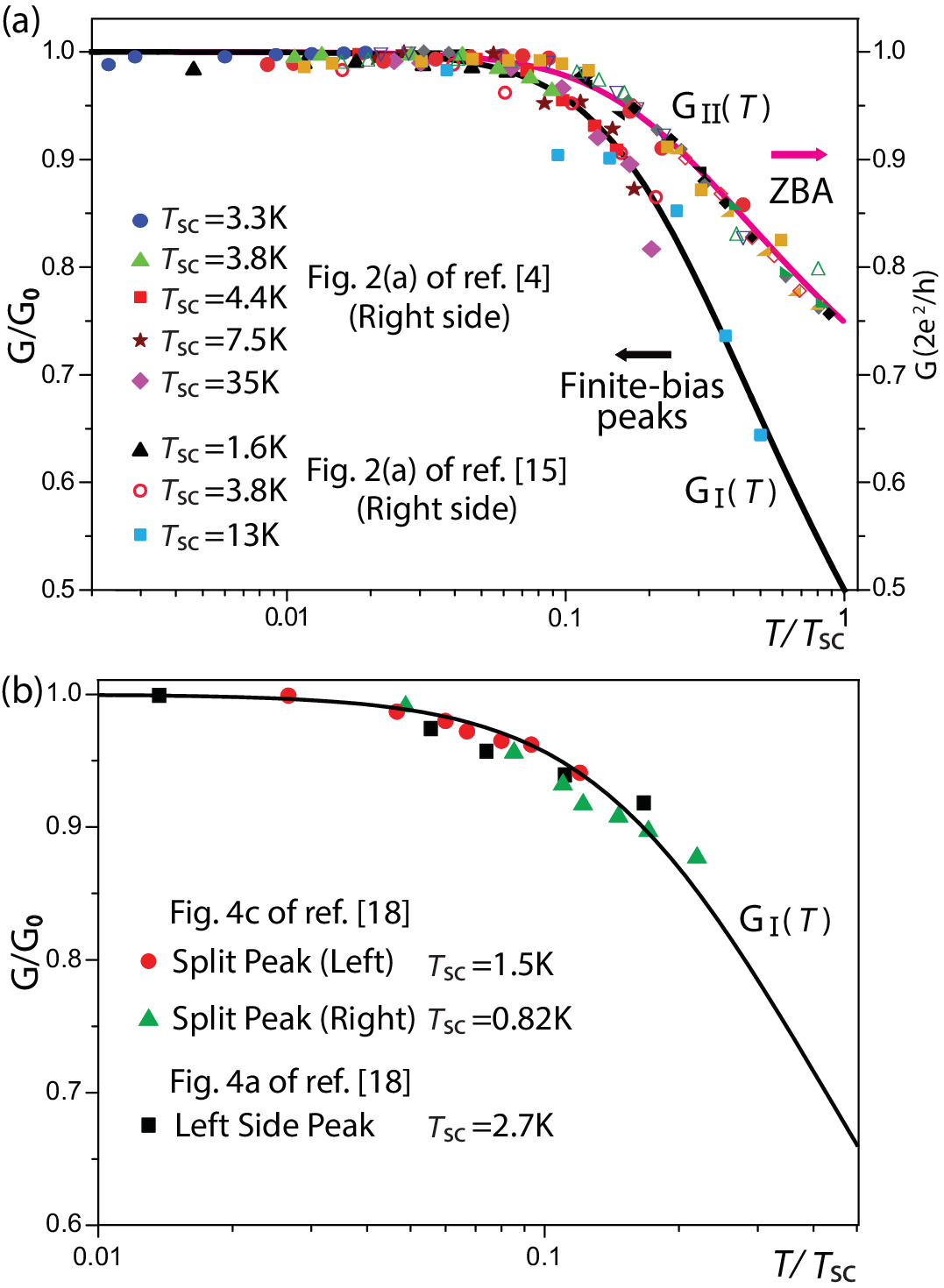}
\caption{
(a) Temperature-scaled finite-bias differential-conductance maxima taken from references~\cite{Cronenwett,Chen} for various gate voltages collapse onto the scaling function $G_{\rm I}(T)$. For comparison, the zero-bias anomaly data from the same references are shown scaled by $G_{\rm II}(T)$. (b) Temperature-scaled finite-bias peaks observed in the even-particle sector of QDSETs collapse onto $G_{\rm I}(T)$. Red dots and green triangles correspond to a spin-singlet state in the QD, whereas black squares correspond to a spin-triplet state~\cite{Roch}. The corresponding values of $T_{\rm SC}$ and the data sources are indicated.
}
\label{fig3}
\end{figure}
%%%%%%%%%%%%%%%%%%%%%%%%%%%%%%%%%%%%%%%%%%%%%%%%%%%%%%%%%%%%%%%%%%%%%%%%%%%%%%%%%%%%%%%%

The reason $G_{\rm 0}$ is used exclusively for the scaling function $G_{\rm I}(T)$ is that the system incorporates a confinement potential essential for quantum dot formation. Since this confinement potential and the subsequent degree of conductance suppression depend on the type of quantum dot, it is appropriate to employ $G_{\rm 0}$ for QDSETs. 
On the other hand, for QPC devices lacking such a confining potential, it is natural to use $G_{\rm II}(T)$, which yields the unitary conductance value of $2e^2/h$ as $T \rightarrow 0$ K.

Because numerous studies have already established that the ZBP in the odd-particle sector of QDSETs follows the scaling function 
$G_{\rm I}(T)$~\cite{Wiel,Heersche,Goldhaber-PRL,Petit}, we focus our scaling analysis on the finite-bias coherent peaks observed in QPC devices~\cite{Cronenwett,Chen} and in the even-particle sectors of QDSETs~\cite{Roch}.
Figure~\ref{fig3}(a) shows that the temperature-dependent finite-bias peaks observed in QPC devices collapse onto the scaling function $G_{\rm I}(T)$. 
For comparison, the same figure also includes the scaling results for the zero-bias anomaly of QPC devices using the scaling function $G_{\rm II}(T)$ reported in references~\cite{Cronenwett,Chen}.
Figure~\ref{fig3}(b) demonstrates that finite-bias peaks appearing in the even-particle sector of QDSETs, irrespective of whether the quantum dot hosts a spin singlet or a spin triplet, also scale according to $G_{\rm I}(T)$. 
As mentioned earlier in this section, Kurzmann $\textit{et al.}$~\cite{Kurzmann} reported similar scaling results for the finite-bias peak observed in a bilayer graphene QDSET exhibiting a spin-triplet state.
These observations indicate that all finite-bias coherent peaks follow the scaling function $G_{\rm I}(T)$, independent of device type.

%%%%%%%%%%%%%%%%%%%%%%%%%%%%%%%%%%%%%%%%%%%%%%%%%%%%%%%%%%%%%%%%%%%%%%%%%%%%%%%%%%%%%%%%
\begin{figure}[t] 
\centering
\includegraphics[width=3.2 in]{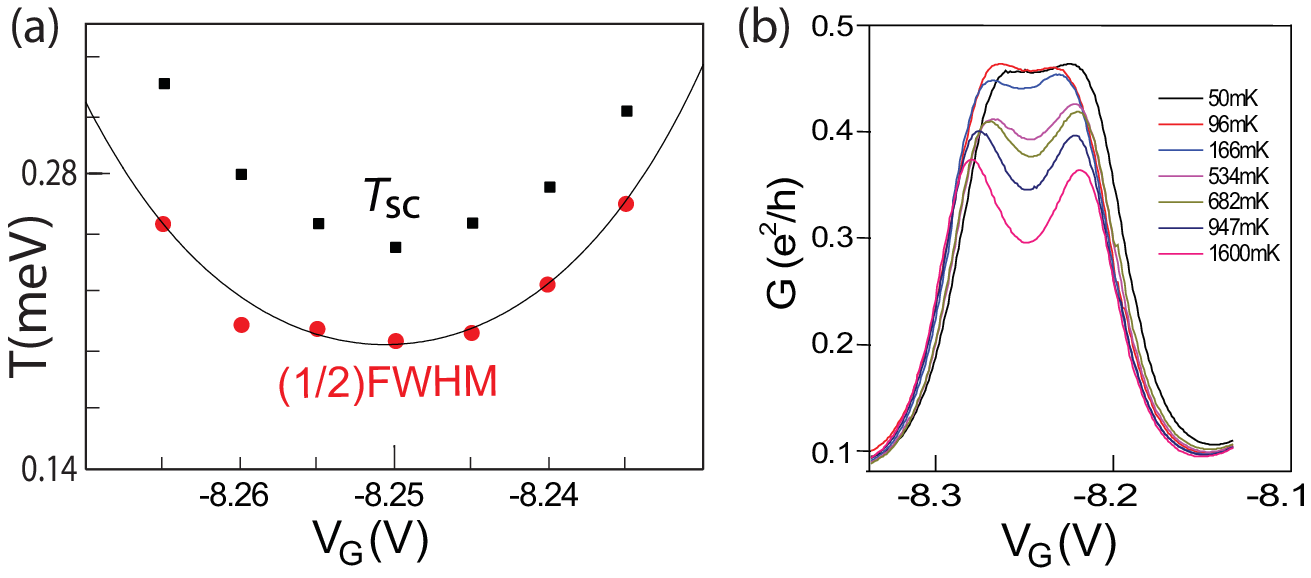}
\caption{(a) Comparison between $T_{\rm SC}$ and $(1/2)$FWHM for a CNT QDSET as a function of $V_{\rm G}$, expressed in the same energy units. The $T_{\rm SC}$ data were obtained through scaling analysis, while the $(1/2)$FWHM data are taken from Fig.~2(d) of reference~\cite{kriss}. (b) Temperature-dependent linear conductance used to determine the $T_{\rm SC}$ values shown in (a).
}
\label{fig4}
\end{figure}
%%%%%%%%%%%%%%%%%%%%%%%%%%%%%%%%%%%%%%%%%%%%%%%%%%%%%%%%%%%%%%%%%%%%%%%%%%%%%%%%%%%%%%%%

It is noteworthy that Cronenwett $\textit{et al.}$~\cite{Cronenwett}, in a study of QPC devices, demonstrated that $T_{\rm SC}$ corresponds to half the full width at half maximum (FWHM) of the ZBP, which represents the Kondo temperature $T_{\rm K}$.
In contrast, van der Wiel $\textit{et al.}$~\cite{Wiel}, investigating the odd-particle sector of a QDSET, showed that the scaling temperature appearing in 
$G_{\rm I}(T)$ does not match half the ZBP FWHM. 
Figure~\ref{fig4}(a) illustrates another instance of this discrepancy observed in the odd-particle sector of a CNT QDSET.
The experimental data shown in figure~\ref{fig4}(b) were provided by the experimental team of reference~\cite{kriss} alongside figure~\ref{fig7} in section~\ref{sec:reproduct}.

An important qualification is that the correspondence between $T_{\rm SC}$ and (1/2)FWHM reported in reference~\cite{Cronenwett} is valid only when the conductance exceeds $0.7(2e^2/h)$.  
Below this value, the singlet-coupling strengths on the left and right sides of the localized spin become unequal, preventing reliable scaling analysis. 
Consequently, the Kondo temperature can be defined only in the regime where the left- and right-singlet couplings are symmetric. 
A detailed discussion is given in reference~\cite{iop-qpc}.

The remaining interesting scaling aspect lies in analyzing the ZBP of a QDSET composed of a spin triplet in its quantum dot~\cite{Kurzmann,Roch}. 
It is clear that $G_{\rm II}(T)$ is a suitable scaling function for the ZBP of QPC devices, and $G_{\rm I}(T)$ is suitable for the ZBP of QDSET odd-particle sectors. 
This raises an interesting question: what is the scaling function governing the ZBP observed in the even-particle sector of the QDSET hosting a spin triplet?

\begin{figure}[b] 
\centering
\includegraphics[width=3.2 in]{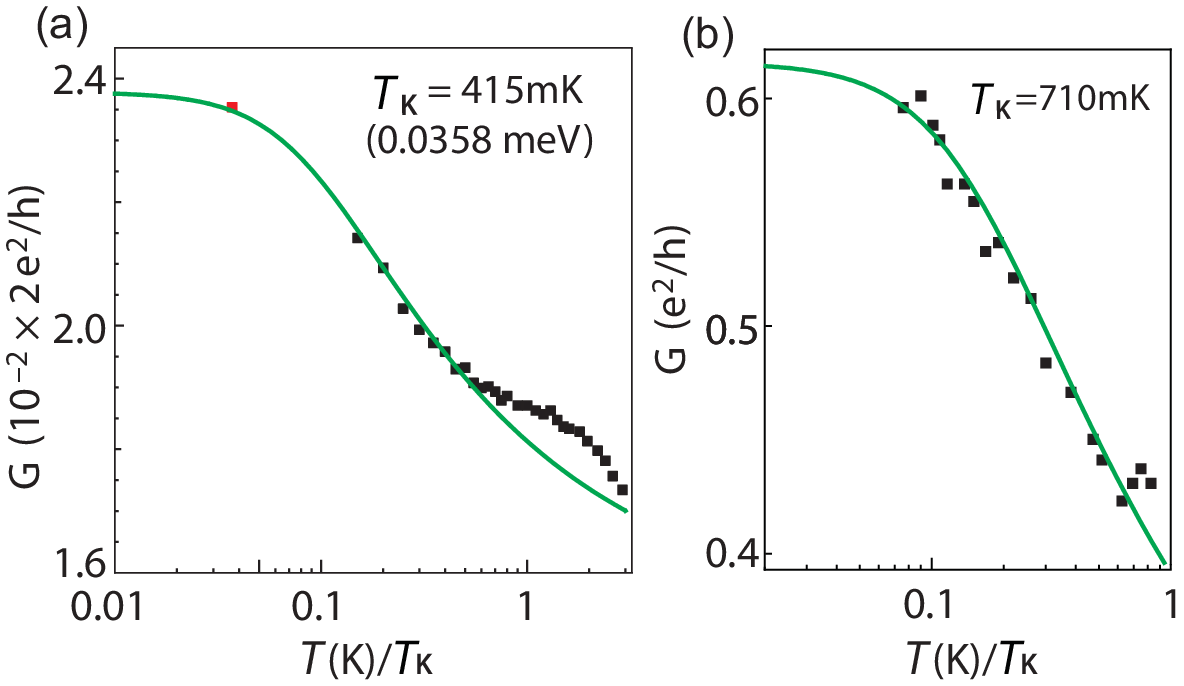}
\caption{
The green curves in panels (a) and (b) represent equations~(\ref{Cond-scale}) and (\ref{Cond-scale2}), respectively. The data shown in (a) and (b) are taken from Fig.~4b of reference~\cite{Roch} and Fig.~3e of reference~\cite{Kurzmann}, respectively. 
The red square in panel (a) denotes the conductance measured at $T=38$ mK, taken from Fig.~4a of reference~\cite{Roch}.
}
\label{fig5}
\end{figure}

Remarkably, as shown in figure~\ref{fig5}, the temperature-dependent ZBP data obtained from a C$_{60}$ QDSET hosting a spin triplet~\cite{Roch} collapse onto the scaling form
\begin{equation} 
G(T)=0.01718\, G_{\rm II}(T)+0.00660\,(2e^2/h)
\label{Cond-scale}
\end{equation}
with scaling temperature $T_{\rm K}=415$ mK (0.0358 meV).
Likewise, the data obtained from the bilayer-graphene QDSET~\cite{Kurzmann} collapse onto
\begin{equation}  
G(T)=0.398\, G_{\rm II}(T)-0.090\,(2e^2/h)
\label{Cond-scale2}
\end{equation}
with scaling temperature $T_{\rm K}=710$ mK.
Here we used the notation $T_{\rm K}$ instead of $T_{\rm SC}$ because the scaling temperature was taken from the (1/2)FWHM of ZBP at the lowest temperature available.

It is noteworthy that the deviations of the scaled data from the scaling functions shown in figure~\ref{fig5} closely resemble those reported in references~\cite{Cronenwett,Wiel,Goldhaber-PRL}, and that the scaling temperatures adopted in references~\cite{Kurzmann,Roch} differ substantially from the corresponding $(1/2)$FWHM values of the lowest-temperature ZBPs.

The coefficient preceding $G_{\rm II}(T)$ plays the role of the sample- and gate-voltage-dependent parameter $G_0$ found in $G_{\rm I}(T)$; this is because the system possesses a confining potential, and the data in figure~\ref{fig5} cannot be scaled using the function $G_{\rm I}(T) + G_{\rm b}$~\cite{Roch}, where $G_{\rm b}$ is the background conductance. 
Therefore, using $G_{\rm II}(T)$ in equations (\ref{Cond-scale}) and (\ref{Cond-scale2}) is the correct choice.

%%%%%%%%%%%%%%%%%%%%%%%%%%%%%%%%%%%%%%%%%%%%%%%%%%%%%%%%%%%%%%%%%%%%%%%%%%%%%%%%%%%%%%%%
%%%%%%%%%%%%%%%%%%%%%%%%%%%%%%%%%%%%%%%%%%%%%%%%%%%%%%%%%%%%%%%%%%%%%%%%%%%%%%%%%%%%%%%%

The limiting values in equation~(\ref{Cond-scale}) are $G(T\rightarrow 0)=2.38\times 10^{-2} (2e^2/h)$ and 
$G(T\rightarrow\infty)=1.52\times 10^{-2} (2e^2/h)$, while those in equation~(\ref{Cond-scale2}) are $G(T\rightarrow 0)=0.308 (2e^2/h)$ and $G(T\rightarrow\infty)=0.055 (2e^2/h)$. 
These limiting values are consistent with the experimental data presented in Fig. 4a of reference~\cite{Roch} and Fig. 3d of reference~\cite{Kurzmann}.
Although high-temperature data are not provided in the latter, the agreement between the limiting values and the experimental data validates the chosen parameters and allows $G(T\rightarrow\infty)$ to be interpreted as the background conductance.

Notably, the small coefficients preceding $G_{\rm II}(T)$ differ by an order of magnitude, despite the systems sharing the same physical components. Since the only difference lies in the materials forming the quantum dots, this suggests that the discrepancy in the coefficients stems from a difference in the coupling function, namely the hybridization strength, between the quantum dot and the reservoir.

The results of the scaling analysis can be summarized as follows:
\begin{enumerate}
\item The ZBP observed in the odd-particle sector of a QDSET follows the scaling function $G_{\rm I}(T)$.

\item All finite-bias coherent peaks, including those observed in the singlet and triplet states of the even-particle sector of a QDSET, also follow $G_{\rm I}(T)$.

\item ZBPs accompanied by two coherent side peaks— namely those observed in QPC devices and in the triplet state of the even-particle sector of a QDSET—follow the scaling function $G_{\rm II}(T)$.
\end{enumerate}

\noindent The scaling analysis demonstrates that the ZBP observed in the odd-particle sector of a QDSET is fundamentally distinct from the ZBP accompanied by coherent side peaks, which appears both in QPC devices and in the even-particle sector of QDSETs that contain a spin triplet in their quantum dots. 
Furthermore, the scaling temperature associated with the former ZBP does not coincide with $(1/2)$FWHM, as shown in figure~\ref{fig4}(a) and reference~\cite{Wiel}, whereas the latter ZBP does. 
The former ZBP therefore 
belongs to the same scaling class as the finite-bias coherent peaks, and its scaling temperature may not be characterized by the conventional Kondo temperature associated with spin exchange.

The next two sections reproduce the gate-voltage-dependent differential-conductance line shapes, demonstrating that the ZBP observed in the odd-particle sector of a QDSET can be understood as the merging of two coherent side peaks and that the spin dynamics responsible for this peak involve only coherent cotunneling of an up--down spin pair intoduced in reference~\cite{Cronenwett98}.

\section{Green's function technique for differential conductance} \label{sec:formula} 

To reproduce the experimentally measured differential-conductance line shapes in the odd-particle sector of a QDSET, we employ the theoretical framework developed in our previous study of QPC devices~\cite{iop-qpc}. 
The gate-voltage-dependent differential conductance, $dI/dV(V_{\rm G})$, is given by~\cite{Meir}:
\begin{equation} 
\frac{dI}{dV}(V_{\rm G}) = \left. \frac{2e^2}{h} \widetilde{\Gamma}^{\rm sn}(V_{\rm G}) {\rm Im} {\mathcal{G}}_{dd\uparrow}^{+\rm sn}(\omega, V_{\rm G}) \right|_{\hbar\omega = eV},
\label{Conductance}
\end{equation}  
where $\widetilde{\Gamma}^{\rm sn}(V_{\rm G})= \Gamma^L(V_{\rm G}) \Gamma^R(V_{\rm G}) / [\Gamma^L(V_{\rm G}) + \Gamma^R(V_{\rm G})]$
denotes the effective $\omega$-independent coupling function of the two-terminal device with the left (right) coupling function $\Gamma^{L(R)}(V_{\rm G})$, and ${\rm Im} \, {\mathcal{G}}_{dd\uparrow}^{+\rm sn}(\omega, V_{\rm G})$ denotes the imaginary part of the on-site retarded Green's function associated with the localized spin state ${\it d}$. 
Meanwhile, the superscript ``sn'' refers to the steady-state nonequilibrium condition. 

Recently, a Green's function approach utilizing the equation-of-motion method has been reported for studying charge and heat transport through quantum dots~\cite{Eckern}. 
While generally useful, this approach has limitations in clearly revealing coherent side peaks and providing detailed information on spin dynamics. The Green's function approach presented in this study provides both of these key pieces of information.

The retarded Green's function $\mathcal{G}_{dd\uparrow}^{+\rm sn}(\omega)$ corresponds to the $dd$ element of the Green-function matrix
\begin{equation} 
i\mathcal{G}_{dd\uparrow}^{+\rm sn}(\omega)= \left[\frac{1}{z \mathbf{I} + i \mathbf{H}}\right]_{dd} = \left[\frac{1}{z \mathbf{I} + i \mathbf{L}}\right]_{dd},
\label{green0}
\end{equation} 
where $z = -i\omega + 0^+$, $\mathbf{I}$ is the identity operator, and $\mathbf{H}$ and $\mathbf{L}$ are the Hamiltonian and Liouville operators, respectively~\cite{Fulde}. 
The second form is derived from many-body Green's functin $\mathcal{G}_{dd\uparrow}^{+}(t)=-i\theta(t)\langle\{d_\uparrow(t),d_\uparrow^\dagger\} \rangle$ in appendix A.

The determination of an appropriate basis set spanning the working space is the most important and technically demanding step in implementing this approach.
For this reason, we formulate the problem in Liouville space rather than Hilbert space, since the former facilitates the systematic construction of basis operators.
The procedure used to determine the basis operators within the two-reservoir Anderson impurity model is outlined in appendix B.

The Hamiltonian is given by:
\begin{eqnarray}
{\bf H}&=&\sum_{\nu\in L,R}\sum_{\sigma,k}[\epsilon^\nu_k c^{\nu\dagger}_{k\sigma} 
c_{k\sigma}^\nu+ \tilde{V}^\nu(V_{\rm G})(d_{\sigma}^{\dagger} c_{k\sigma}^\nu+c_{k\sigma}^{\dagger\nu} d_{\sigma})] \nonumber \\
&+&\sum_{\sigma} \epsilon_d(V_{\rm G}) d^{\dagger}_{\sigma}d_{\sigma}+ U(V_{\rm G}) n_{d\uparrow} n_{d\downarrow},
\label{Hamiltonian}
\end{eqnarray}
where $c_{k\sigma}^\nu$ ($c_{k\sigma}^{\nu\dagger}$) denotes the annihilation (creation) operator of an electron with spin $\sigma$ and quantum number $k$ in reservoir $\nu$, $\epsilon_k$ is the corresponding kinetic energy, and $n_{d\uparrow}=d_\uparrow^\dagger d_\uparrow$ is the up-spin number operator. The quantities $\tilde V^\nu(V_{\rm G})$ and $\epsilon_d(V_{\rm G})$ denote the gate-voltage-dependent hybridization strength and the energy of the localized $d$ level, respectively. For simplicity, the hybridization is assumed to be real and independent of $k$.
To account for the Coulomb-blockade effect in the quantum dot, an explicit gate-voltage dependence is introduced into $U(V_{\rm G})$.

According to the results obtained in appendix B, the working Liouville space is spanned by the following two groups of basis operators:
\begin{eqnarray*}
{\rm I}:&\{& \delta j^{\pm L}_{d\downarrow}d_{\uparrow}, d_{\uparrow},  \delta j^{\pm R}_{d\downarrow}d_{\uparrow}\}, \\
{\rm II}:&\{& c_{k\uparrow}^{L}, \delta n_{d\downarrow} c_{k\uparrow}^{L}, \delta n_{d\downarrow} c_{k\uparrow}^{R}, c_{k\uparrow}^{R} \} \, \mbox{with} \, k = 0, 1, \cdots, \infty,
\end{eqnarray*}
where $\delta$ denotes the fluctuation of an operator about its expectation value,
$j^-_{d\downarrow}=i\sum_k{\tilde{V}}(d^\dagger_\downarrow c_{k\downarrow}-c^\dagger_{k\downarrow}d_\downarrow)$, and 
$j^+_{d\downarrow}=\sum_k{\tilde{V}}(d_\downarrow^\dagger c_{k\downarrow}+c^\dagger_{k\downarrow}d_\downarrow)$.
The inner product in the Liouville space is defined as $({\widehat A},{\widehat B}):=\langle {\widehat A}{\widehat B}^\dagger+{\widehat B}^\dagger{\widehat A}\rangle$, where ${\widehat B}^\dagger$ is the adjoint of ${\widehat B}$ and the angular brackets denote the expectation.

\begin{figure}[t] 
\centering
\includegraphics[width=2.8 in]{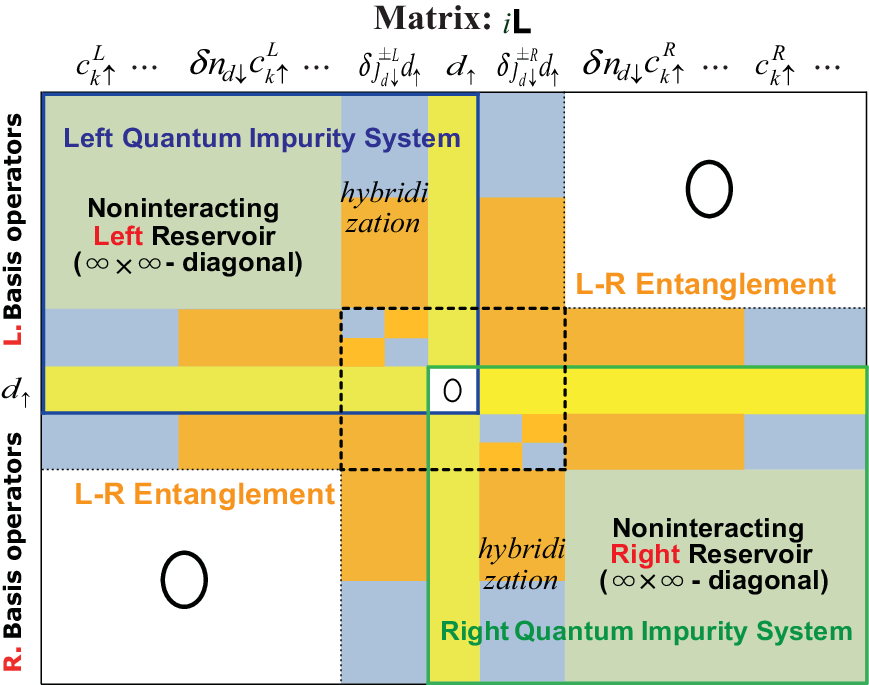}
\caption{
Schematic structure of the Liouville matrix. The different sectors correspond to noninteracting reservoirs (green), vanishing left--right entanglement (cyan), nonvanishing left--right entanglement (orange), superposition between $d_\uparrow$ and other basis operators (yellow), and prohibited direct hopping between reservoirs (white).
}
\label{fig6}
\end{figure}

Using these basis operators arranged as shown at the top of figure~\ref{fig6}, we construct an infinite-dimensional Liouville matrix $i{\bf L}$ by 
calculating matrix elements.
The calculation procedures of matrix elements are detailed in appendix C. 
The structure of Liouville matrix $i{\bf L}$ is schematically shown in figure~\ref{fig6} and detailed in figure~\ref{figD1} in appendix D. 

The infinite-dimensional matrix ${\rm\bf M}$,  where ${\rm\bf M}=z{\bf I}+i{\bf L}$, can be reduced to an equivalent $5 \times 5$ matrix $\mathbf{M}_r$ using the matrix-reduction scheme developed in references~\cite{Lowdin,Mujica}.
The matrix-reduction procedure described in appendix D yields the reduced matrix $\mathbf{M}_r$ as follows:
\begin{eqnarray}
{\rm\bf M}_r=\left(\begin{array}{c c c c c} -i\tilde{\omega} & \gamma^L &
-U^L_{j^-} & \gamma^{LR}_S & \gamma^{LR}_A \\ -\gamma^L & -i\tilde{\omega}
& -U^L_{j^+} & \gamma^{LR}_A & \gamma^{LR}_S \\
U_{j^-}^{L*} &  U_{j^+}^{L*} & -i\tilde{\omega} &  U^{R*}_{j^+} &
U^{R*}_{j^-} \\  -\gamma^{LR}_S & -\gamma^{LR}_A & -U_{j^+}^R  &
-i\tilde{\omega} & -\gamma^R \\
 -\gamma^{LR}_A &  -\gamma^{LR}_S &  -U_{j^-}^R  & \gamma^R & -i\tilde{\omega}
\end{array}\right)+i{\bf \Sigma}, \nonumber \\
\label{reduced}
\end{eqnarray} 
where $\tilde{\omega}=\omega-\epsilon_d-U\langle n_{d\downarrow}\rangle$.
The matrix elements of the first term derived in appendix C are given by

%%%%%%%%%%%%%%%%%%%%%%%%%%%%%%%%%%%%%%%%%%%%%%%%%%%%%%%%%%%%%%%%%%%%%%%%%%%%%%%%%%%%%%%%
%%%%%%%%%%%%%%%%%%%%%%%%%%%%%%%%%%%%%%%%%%%%%%%%%%%%%%%%%%%%%%%%%%%%%%%%%%%%%%%%%%%%%%%%

\begin{eqnarray}
\gamma^{L(R)}=\frac{\langle\sum_ki(\tilde{V}^Lc_{k\uparrow}^L+\tilde{V}^Rc^R_{k\uparrow})d^\dagger_{\uparrow}
		[j^{-L(R)}_{d\downarrow},j^{+L(R)}_{d\downarrow}]\rangle^{\rm sn}}
{\langle(\delta j^{-L(R)}_{d\downarrow})^2\rangle^{1/2}\langle(\delta j^{+L(R)}_{d\downarrow})^2\rangle^{1/2}}, \nonumber \\
\label{gamma-L}
\end{eqnarray}
\begin{equation}
\gamma^{LR}_S=\frac{\langle\sum_ki(\tilde{V}^Lc_{k\uparrow}^L+\tilde{V}^Rc^R_{k\uparrow})d^\dagger_{\uparrow}
		[j^{-L}_{d\downarrow},j^{+R}_{d\downarrow}]\rangle^{\rm sn}}
{\langle(\delta j^{-L}_{d\downarrow})^2\rangle^{1/2}\langle(\delta j^{+R}_{d\downarrow})^2\rangle^{1/2}}, 
\label{gamma-S}
\end{equation}
\begin{equation}
\gamma^{LR}_A=\frac{\langle\sum_ki(\tilde{V}^Lc_{k\uparrow}^L+\tilde{V}^Rc^R_{k\uparrow})d^\dagger_{\uparrow}
		[j^{-L}_{d\downarrow},j^{-R}_{d\downarrow}]\rangle^{\rm sn}}
{\langle(\delta j^{-L}_{d\downarrow})^2\rangle^{1/2}\langle(\delta j^{-R}_{d\downarrow})^2\rangle^{1/2}}, 
\label{gamma-A}
\end{equation}  
and
\begin{eqnarray}
U_{j^\mp}^{\nu}=i\frac{U}{2}\frac{\langle [n_{d\downarrow},j^{\mp\nu}_{d\downarrow}]
(1-2n_{d\uparrow})+j^{\mp\nu}_{d\downarrow}(1-2\langle n_{d\downarrow}\rangle)\rangle^{\rm eq}}
{\langle(\delta j^{\mp\nu}_{d\downarrow})^2\rangle^{1/2}}, \nonumber \\
\label{U} 
\end{eqnarray}
where the superscript `eq' in \(U_{j^\mp}^{\nu}\) indicates that the expectation is taken at equilibrium.

The dynamical processes encoded in the matrix elements $\gamma^L$ and $\gamma^R$ describe unidirectional entangled-state tunneling accompanied by spin exchange, and those encoded in $\gamma^{LR}_{S,A}$ involve singlet cotunneling only, as illustrated in the upper and lower panels of figure~\ref{fig2}, respectively.
A detailed description is given in appendix E.
In contrast, the parameters $U_{j^\mp}^{L,R}$ represent effective gate-voltage-dependent Coulomb interactions that affect the motion of incoherent electrons flowing to or from the left and right reservoirs.
In QDSETs, the gate-voltage dependence of $U_{j^\mp}^{L,R}$ reflects the Coulomb-blockade effect. 

The second term in equation~(\ref{reduced}), namely $i{\bf \Sigma}$, represents the self-energy matrix generated during the matrix-reduction procedure shown in appendix D. 
Its matrix elements are given by
$i{\bf \Sigma}_{pq}=\eta_{pq}\left[i{\bf \Sigma}_0^L(\omega)+i{\bf \Sigma}_0^R(\omega)\right]$,
where the coefficients $\eta_{pq}$ arise naturally from the reduction procedure and $i{\bf \Sigma}_0^\nu(\omega)=\pi(\tilde V^\nu)^2\rho_0 (\sqrt{1-\omega^2/D^2}+i\omega/D)$ is the self-energy of the noninteracting Anderson model with a wide semi-elliptic band,
$\rho_0(\omega)=\rho_0\sqrt{1-\omega^2/D^2}$.
The expression for the coefficient $\eta_{pq}$ is not simple, as it incorporates the effects of left--right entanglement. 
Appendix F details the specific expressions for $\eta_{pq}$, the standard value of 1/4 given in the absence of left--right entanglement, and the inequality 
$\eta_{11}=\eta_{15}=\eta_{55}<\eta_{12}=\eta_{14}=\eta_{25}=\eta_{45}<\eta_{22}=\eta_{24}=\eta_{44}$ that arises from the contribution of these entanglement effects.

Meanwhile,  the coupling function $\Gamma^\nu(\omega)$ appearing in the differential conductance corresponds to twice the imaginary part of ${\bf \Sigma}_0^\nu(\omega)$, i.e., $\Gamma^\nu(\omega)=2\pi(\tilde{V}^\nu)^2\rho_0(\omega)$.
Assuming a sufficiently wide bandwidth, $\Gamma^\nu(\omega)$ can be treated as a constant with respect to $\omega$, denoted by  
$\Gamma^\nu=2\pi(\tilde{V}^\nu)^2\rho_0=2\Delta^\nu$.
The parameter $\Delta^\nu$ depends on gate voltage $V_{\rm G}$ through the hybridization $\tilde{V}^\nu(V_{\rm G})$, while the sum $\Delta^L + \Delta^R$ remains constant. 
Throughout this work, energies are measured in units of $\Delta=(\Delta^L + \Delta^R)/2$.

Consequently, the relation $-{\rm Im}{\mathcal G}^{+}_{dd\uparrow}(\omega)={\rm Re}({\rm\bf M}_r^{-1})_{33}$, given by equation (\ref{eq:M-matrix3})  derived in detail in appendix D, provides the following compact expression for the differential conductance:
\begin{equation}
\frac{dI}{dV}= \left. \frac{2e^2}{h} \widetilde{\Gamma}^{\rm sn}(V_{\rm G}){\rm Re}({\rm\bf M}_r^{-1})_{33}\right|_{\hbar\tilde{\omega}=eV}.
\label{eq:new-dIdV}
\end{equation}
This expression serves as the basis for reproducing the experimentally observed differential-conductance line shapes presented in the next section.
Since a wide flat band and the relation $\epsilon_d\approx -U\langle n_{d\downarrow}\rangle$ are assumed, the difference between $\tilde{\omega}$ 
and $\omega$ is not an issue.

\section{Theoretical reproduction of differential- conductance line shapes} \label{sec:reproduct} 

The gate-voltage-dependent differential conductance of the CNT QDSET was measured in the range of $-9.3 {\rm V}\leq V_{\rm G}\leq -7.8 {\rm V}$; most of these results are presented as a grayscale plot in Fig.~1(b) of reference~\cite{kriss}, with several $dI/dV$ line shapes superimposed.
All measurements were performed using a dilution refrigerator at a base temperature below 100 mK.
An external magnetic field of 1 kG was applied to suppress the superconductivity of the aluminum electrodes and drive the system into the normal state.

%%%%%%%%%%%%%%%%%%%%%%%%%%%%%%%%%%%%%%%%%%%%%%%%%%%%%%%%%%%%%%%%%%%%%%%%%%%%%%%%%%%%%%%%
\begin{figure}[b] 
\centering
\includegraphics[width=3.2 in]{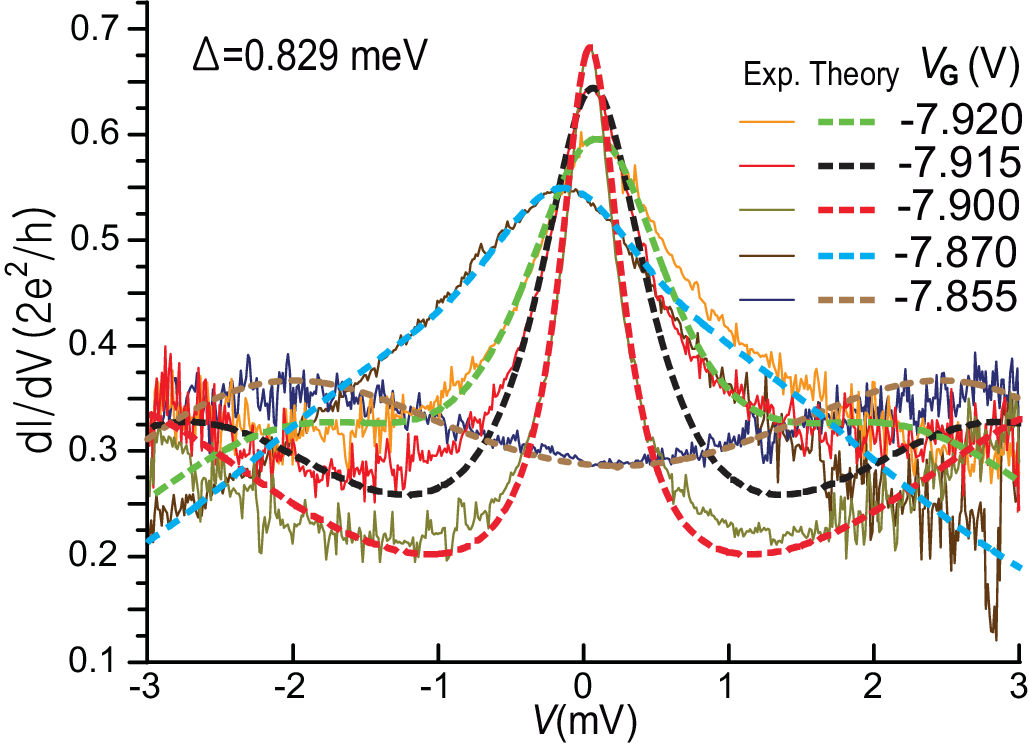}
\caption{
Experimental differential-conductance curves (thin solid lines) for a CNT QDSET and the corresponding theoretical line shapes (thick dashed lines) for various gate voltages. The theoretical curves have been shifted slightly to align with the experimental peak positions. The spectrum at $V_{\rm G}=-7.855$ V, which does not exhibit a ZBP, lies outside the odd-particle sector of the Coulomb diamond.
}
\label{fig7}
\end{figure}
%%%%%%%%%%%%%%%%%%%%%%%%%%%%%%%%%%%%%%%%%%%%%%%%%%%%%%%%%%%%%%%%%%%%%%%%%%%%%%%%%%%%%%%%

The unpublished differential conductance data for the range $-7.920 {\rm V}\leq V_{\rm G}\leq -7.855 {\rm V}$, shown in figure~\ref{fig7}, were obtained through the cooperation of the relevant experimental group for this study.
The measured differential-conductance spectra exhibit a central ZBP accompanied by two broad Coulomb-blockade peaks, a characteristic feature of the odd-particle sector of a QDSET~\cite{Nygard,Wiel,Heersche}. 

The experimental line shapes are reproduced theoretically by determining the parameters appearing in equation~(\ref{reduced}),  together with the effective coupling factor $\widetilde{\Gamma}^{\rm sn}(V_{\rm G})$ entering equation~(\ref{eq:new-dIdV}).
To this end, analysis in the atomic limit (${\rm Im}\Sigma(\omega) \rightarrow 0$) in the large-$U$ limit at half-filling (${\rm Im}[U_{j^\pm}^{L,R}]=0$)
provides the following important insight~\cite{Hong11}:
\begin{enumerate}
\item The spectral weight of the ZBP is controlled by
$$\left[1+\frac{U^2\{(\gamma^L)^2+(\gamma^R)^2+2(\gamma_S^{LR}-\gamma_A^{LR})^2\}}{8\{\gamma^L\gamma^R+(\gamma_S^{LR})^2-(\gamma_A^{LR})^2\}^2}\right]^{-1}.$$
\item The positions of the coherent side peaks are given by
$\pm\sqrt{\{(\gamma^L)^2+(\gamma^R)^2\}/2+(\gamma_S^{LR}-\gamma_A^{LR})^2}$.
\item The spectral weight of the coherent side peaks for $\gamma^L=\gamma^R$ is determined by
$$\frac{8(\gamma^L)^2(\gamma_A^{LR})^2}{U^2\left[(\gamma^L)^2+(\gamma_S^{LR}-\gamma_A^{LR})^2\right]}.$$
\end{enumerate}  

Appendix E shows that $\gamma^{LR}_{S}$ and $\gamma^{LR}_{A}$ represent a symmetric and antisymmetric combinations of the left-right flows between the two reservoirs, respectively.
At non-zero bias, entangled-state tunneling becomes unidirectional, resulting in $\gamma^{LR}_{A}=\gamma^{LR}_{S}$. 
Consequently, the spectral weight of ZBP, relation (i), becomes $4(\gamma^L)^2/U^2$ and the position of the coherent side peak, relation (ii), becomes 
$\pm\gamma^L$.
Relation (iii) carries significant meaning. This is discussed in the final paragraph of this section.

These observations suggest that the condition $\gamma^L=\gamma^R=0$ is appropriate for the evolution of three coherence peaks, characteristic of QPCs, into the single ZBP observed in the odd-particle sector of a QDSET.
In other words, this choice suppresses the zero-bias anomaly associated with spin exchange in QPC devices according to relation (i), while relation (ii) simultaneously drives the coherent side peaks toward zero bias. 
A smooth single ZBP is obtained when $U_{j^-}^{L} = U_{j^+}^{L} = U_{j^-}^{R} = U_{j^+}^{R}$.
This condition is discussed in detail in the following section.
Thus, only two independent parameters, namely $\gamma^{LR}_{S,A}$ controlling the ZBP spectral weight and $U_{j^\mp}^{L,R}$ controlling the position of Coulomb-blockade peak, remain except the self-energy coefficients $\eta_{pq}$ and the prefactor $\widetilde{\Gamma}^{\rm sn}(V_{\rm G})$ in equation (\ref{eq:new-dIdV}).

The self-energy coefficients $\eta_{pq}$ are determined phenomenologically based on the inequality presented in the preceding section; this inequality accounts for the standard value of $0.25$ derived in the absence of entanglement effects and the contribution of entanglement described in appendices F and C, respectively. 
Thus, we adopt 
$\eta_{11} = \eta_{15} = \eta_{55} = 0.253$, 
$\eta_{12} = \eta_{14} = \eta_{25} = \eta_{45} = 0.254$, 
$\eta_{22} = \eta_{24} = \eta_{44} = 0.260$, and 
$\eta_{33} = 1$, with the remaining coefficients obtained by symmetry.
The validity of these values has been verified by their successful application in studies of QPC devices utilizing the same Hamiltonian~\cite{iop-qpc}.
In contrast, $\widetilde{\Gamma}^{\rm sn}(V_{\rm G})$ can be independently determined by matching the theoretical and experimental peak amplitudes.

%%%%%%%%%%%%%%%%%%%%%%%%%%%%%%%%%%%%%%%%%%%%%%%%%%%%%%%%%%%%%%%%%%%%%%%%%%%%%%%%%%%%%%%%
\begin{table} [t]
\centering
\caption{\textbf{Gate-voltage dependence of the fitting parameters.}}
   \vspace{0.3cm}
    \setlength{\tabcolsep}{2.7 pt}
   \begin{tabular}{c c c c c c}
\hline\hline \\ [-2ex] $V_{\rm G}$ & $\gamma^L$ & $\gamma^R$ & $\gamma^{LR}_{S,A}$ & $U_{J^\pm}^{L,R}$ & 
$\widetilde{\Gamma}^{\rm sn}$
\\ [0.5ex] \hline  
-7.920 & 0   & 0     &  0.72   & 0.72     &  0.5955    \\
-7.915 & 0   & 0     &  0.75   & 1.10     &  0.6445    \\
-7.900 & 0   & 0     &  0.67   & 1.48     &  0.6800    \\
-7.870 & 0   & 0     &  0.60   & 0.34     &  0.5480   \\
-7.855 & 0   & 0     &  0.00   & 1.10     &  0.6250    \\
 [0.5ex] \hline
\end{tabular}
\label{table}
\end{table}
%%%%%%%%%%%%%%%%%%%%%%%%%%%%%%%%%%%%%%%%%%%%%%%%%%%%%%%%%%%%%%%%%%%%%%%%%%%%%%%%%%%%%%%%

The theoretical curves calculated using the parameters listed in table~\ref{table} nicely fit the experimental data after aligning the peak positions, as shown in figure~\ref{fig7}. 
Table~\ref{table} and figure~\ref{fig7} show that $\gamma_{S}^{LR}=\gamma_{A}^{LR}=0$ at $V_{\rm G}=-7.855$ V gives rise to the complete disappearance of the coherent side peaks responsible for the ZBP, leaving only the Coulomb-blockade peaks.
Furthermore, the gate-voltage dependence trend of $U_{j^\mp}^{L,R}$ shown in table~\ref{table} indicates that $V_{\rm G}=-7.855$ V corresponds to the even-particle sector adjacent to the odd-particle sector to which the other gate voltages in table~\ref{table} belong. 
This suggests that the gate-voltage dependence of $U_{j^\mp}^{L,R}$ is consistent with the evolution of the Coulomb-diamond boundary.

We conclude this section with several remarks regarding the appearance and disappearance of the coherent side peaks. 
Appendix E shows that $\gamma^{LR}_{A}$ represents an antisymmetric combination of the left-right flows between the two reservoirs and thus vanishes in equilibrium and has a finite value under bias. 
Accordingly, relation (iii) above implies that the coherent side peaks disappear in equilibrium and appear under bias. 
This observation indicates that the coherent side peaks originate not from standard resonant tunneling through the quantum dot's energy levels, but from coherent cotunneling driven by many-body correlations under nonequilibrium steady-state conditions.
This is precisely why scaling can be applied to the finite-bias peaks, as discussed in section~\ref{sec:scaling}.

\section{Discussion} \label{sec:discuss} 

We first examine whether the two conditions imposed in reproducing the differential-conductance line shapes in the previous section, namely $\gamma^{L,R}=0$ and $U_{j^-}^{L}=U_{j^+}^{L}=U_{j^-}^{R}=U_{j^+}^{R}$, are mutually consistent.

To address this question, we note that the backward component of electron motion appears only during the spin-exchange processes encoded in $\gamma^L$ and $\gamma^R$, as illustrated in the upper panel of figure~\ref{fig2}. 
Suppressing backward motion therefore naturally leads to the condition $\gamma^L=\gamma^R=0$.
By contrast, the parameters $U_{j^\mp}^{L,R}$ characterize effective Coulomb interactions that influence the motion of incoherent electrons between the reservoirs and the localized spin. 
The subscripts $j^-$ and $j^+$ denote the corresponding modes of electron motion between the reservoir and the localized spin, namely $\longrightarrow-\longleftarrow$ and $\longrightarrow+\longleftarrow$, respectively. 
Assuming a symmetric Coulomb blockade effect, requiring the absence of backward motion therefore leads to the condition $U_{j^-}^{L}=U_{j^+}^{L}=U_{j^-}^{R}=U_{j^+}^{R}$.
The two conditions thus originate from the same physical requirement and are therefore mutually consistent.

The most important implication of figure~\ref{fig7} is that the ZBP observed in the odd-particle sector of a QDSET can be understood not as a coherent resonance formed by spin dynamics involving spin exchange and cotunneling—such as the zero-bias anomaly in QPC devices—but rather as the merging of two coherent side peaks generated solely by singlet cotunneling.
Within the present theoretical framework, these coherent side peaks of QDSET correspond to the explicitly separated side peaks in QPC devices~\cite{Kristensen,Cronenwett,Chen,Sarkozy,Ren,DiCarlo}. 
This interpretation is consistent with the scaling results presented in section~\ref{sec:scaling}, where finite-bias coherent peaks and the ZBP of the odd-particle sector of a QDSET were found to belong to the same scaling class.

More generally, the present analysis suggests that coherent transport in nanodevices under bias is governed fundamentally by singlet cotunneling~\cite{Cronenwett98}, whereas spin exchange provides an additional dynamical process that distinguishes a specific class of coherent peaks. 
From this perspective, the classification proposed in this work extends and refines the microscopic understanding of coherent transport phenomena occurring in nanoelectronic devices under steady-state nonequilibrium based on the two-reservoir Anderson impurity model.

\section{Conclusion}\label{sec:conclusion}

We have performed a scaling analysis of temperature-dependent coherent peaks observed in QDSETs and QPC devices and demonstrated that these peaks can be classified into two categories according to both their scaling behavior and their underlying spin dynamics.
The first category comprises the ZBP observed in the odd-particle sector of a QDSET together with all finite-bias coherent peaks. 
These peaks follow the scaling function $G_{\rm I}(T)$ of equation (\ref{SF-1}) and are associated with transport processes governed by coherent cotunneling of an up--down spin pair~\cite{Cronenwett98}, which is referred to in this study as singlet cotunneling.
The second category consists of the ZBPs observed in the even-particle sector of QDSETs containing a spin triplet within the quantum dot and the ZBPs observed in QPC devices.
These peaks follow the scaling function  $G_{\rm II}(T)$ of equation (\ref{SF-2}) and involve spin-exchange dynamics in addition to singlet cotunneling.
Furthermore, the theoretical reproduction of the differential-conductance line shapes indicates that the ZBP observed in the odd-particle sector of a QDSET can be understood as the merging of two coherent side peaks generated by singlet cotunneling.
This interpretation consistently explains both the observed scaling behavior and the relationship between the scaling temperature and the zero-bias peak width.

In this picture, singlet cotunneling constitutes the fundamental transport process responsible for coherent current flow, whereas spin exchange provides an additional dynamical contribution that manifests itself in specific classes of coherent peaks. The present classification therefore offers a more detailed microscopic framework for understanding coherent transport phenomena in QDSET and QPC devices.

%\newpage

\centerline{\bf ACKNOWLEDGMENTS}

The author appreciates B.-K. Kim and M.-H. Bae of the Korea Research Institute of Standards and Science for providing unpublished data on differential conductance.

\appendix   

\section{Green's function in the Liouville space}

\setcounter{section}{1}

The second form of the retarded Green's function in equation~(\ref{green0}) can be derived from the usual on-site many-body retarded Green's function in time space,
$\mathcal{G}_{dd\uparrow}^{+}(t)=-i\theta(t)\langle\{d_\uparrow(t),d_\uparrow^\dagger\} \rangle$,
where the curly brackets denote the anticommutator and the angular brackets indicate
an average. Now, we consider an expansion
$d_\uparrow(t)=\sum_{i=1}^\infty A_i(t)\hat{e}_i$
in the Liouville space in which the inner product is defined as
$(\hat{e}_i,\hat{e}_j):=\langle\{\hat{e}_i,\hat{e}_j^\dagger\}\rangle$,
where $\hat{e}_1=d_\uparrow$ and the set of orthonormal vectors
$\{\hat{e}_j | j = 1, 2, \cdots ,\infty\}$ is complete.

We apply the identity $(d/dt)=i{\bf L}$, where ${\bf L}\hat{e}_i\equiv[{\bf H},\hat{e}_i]$
and ${\bf H}$ is the Hamiltonian, to the summation $\sum_{i=1}^\infty A_i(t)\hat{e}_i$.
Then, we have the relationship:
\begin{equation}
\sum_{i=1}^\infty \frac{dA_i(t)}{dt}\hat{e}_i
=\sum_{i=1}^\infty A_i(t)i{\bf L}\hat{e}_i.
\label{eq:GR01}
\end{equation}
Taking inner products with $\hat{e}_j$ for both sides leads to
\begin{equation}
\frac{dA_j(t)}{dt}=\sum_{i=1}^\infty \langle i{\bf L}\hat{e}_i|\hat{e}_j\rangle A_i(t).
\label{eq:GR02}
\end{equation}
Laplace transform of equation (\ref{eq:GR02}) using the boundary conditions $A_1(t = 0)=1$ and $A_j(t=0)=0$ for $j\neq 1$
is written as
\begin{equation}
zA_j(z)- A_j(t=0)=\sum_{i=1}^\infty m_{ji}A_i(z),
\label{eq:GR03}
\end{equation}
where $z=-i\omega+0^+$ and $m_{ji}=(i{\bf L}\hat{e}_i,\hat{e}_j)=-(\hat{e}_i,i{\bf L}\hat{e}_j)$,
which is guaranteed when the inner product has time-translational invariance. 
Since we consider an equilibrium or steady-state nonequilibrium, this relation is satisfied.
Note that $(\hat{e}_i,i{\bf L}\hat{e}_j)$ denotes the matrix elements of the Liouville matrix $i{\bf L}$.

Matrix representation of equation (\ref{eq:GR03}) is written as
$(z{\bf I}-{\bf m}){\bf A}_c=(z{\bf I}+i{\bf L}){\bf A}_c={\bf D}_c$, where the symbol ${\bf I}$ denotes
the identity operator, and ${\bf A}_c$ and ${\bf D}_c$ are column matrices
defined by ${\bf A}^T_c=(A_1(z),\cdots,A_\infty(z))$ and
${\bf D}^T_c=(1, 0,\cdots,0)$ due to the boundary condition given above, respectively.
Here, the superscript $T$ denotes the transpose. 
Note that $A_1(z)$, namely the Laplace transform of $A_1(t)=\langle\{d_\uparrow(t),d_\uparrow^\dagger\}\rangle$,
yields the on-site retarded Green's function $i\mathcal{G}_{dd\uparrow}^{+}(\omega)$.

Since ${\bf A}_c=(z{\bf I}+i{\bf L})^{-1}{\bf D}_c$, the first element of ${\bf A}_c$, i.e., $A_1(z)$, is given by
\begin{equation}
A_1(z)=i\mathcal{G}_{dd\uparrow}^{+}(z)=({\bf M}^{-1})_{11},
\label{eq:GR04}
\end{equation}
where ${\bf M}=(z{\bf I}+i{\bf L})$.
Thus, the second resolvent operator form of equation~(\ref{green0}) is obtained.

\section{Determination of basis operators}

\setcounter{section}{2}

We begin with the retarded Green’s function $i\mathcal{G}_{dd\uparrow}^{+}(z)$ given in equation~(\ref{green0}) or equation (\ref{eq:GR04}), which is the $dd$ element of the Green’s function operator $\widehat{G}$ represented by a matrix:
\begin{equation} 
i\widehat{G} = \frac{1}{z \mathbf{I} + i \mathbf{L}}.
\label{green}
\end{equation} 

A systematic method to determine a complete set of basis operators spanning the Liouville space of $d_\sigma(t)$ begins with separating the Liouville operator into two parts: ${\bf L}={\bf L}_I+{\bf L}_h$, where ${\bf L}_I$ and ${\bf L}_h$ represent the Liouville operators using the isolated part of the Hamiltonian 
${\bf H}_I={\bf H}-{\bf H}_h$ and the hybridization part ${\bf H}_h$ in equation (\ref{Hamiltonian}). 
Subsequently, we set ${\widehat A}=z{\bf I}/2+i{\bf L}_I$ and ${\widehat B}=z{\bf I}/2+i{\bf L}_h$ and utilize an operator identity:
\begin{eqnarray}
({\widehat A}+{\widehat B})^{-1} = {\widehat A}^{-1}-{\widehat A}^{-1}{\widehat B}{\widehat A}^{-1}+({\widehat A}^{-1}{\widehat B}){\widehat A}^{-1}({\widehat B}{\widehat A}^{-1}) \nonumber \\
-({\widehat A}^{-1}{\widehat B}){\widehat A}^{-1}{\widehat B}{\widehat A}^{-1}({\widehat B}{\widehat A}^{-1}) +\cdots  \label{opexpan} \\
={\widehat A}^{-1}-{\widehat A}^{-1}{\widehat B}{\widehat A}^{-1}+({\widehat A}^{-1}{\widehat B})({\widehat A}+{\widehat B})^{-1}({\widehat B}{\widehat A}^{-1}). \, \,
\label{opexpan2}
\end{eqnarray}

Equation (\ref{opexpan2}) writes the retarded Green’s function $i\mathcal{G}_{ij\sigma}^+(z)=\langle e_{i\sigma}| i\widehat{G}|e_{j\sigma}\rangle$ as \begin{eqnarray}
i\mathcal{G}_{ij\sigma}^+(z)&=&\langle e_{i\sigma}| i\widehat{G}_{I}|e_{j\sigma}\rangle -\langle e_{i\sigma}| i\widehat{G}_{I}\, {\widehat B}\, i\widehat{G}_{I}| e_{j\sigma}\rangle \nonumber \\
&+&\langle e_{i\sigma}| i\widehat{G}_{I}\, {\widehat B}\, i\widehat{G}\, {\widehat B}\, i\widehat{G}_{I}|e_{j\sigma}\rangle, 
\label{expand3}
\end{eqnarray}
where the ket $|e_{j\sigma}\rangle\equiv {\hat e}_{j\sigma}|0 \rangle$ with ${\hat e}_{j\sigma}$ denoting a basis operator of a complete set 
$\{ {\hat e}_{j\sigma}\}$, $i=1, 2, \cdots, \infty$, and $|0\rangle$ the ground state.

The first term in equation (\ref{expand3}), $\langle e_{i\sigma}| i{\widehat G}_{I}|e_{j\sigma}\rangle$, represents the matrix element $(i{\widehat G}_I)_{ij}=[(z{\bf I}/2+i{\bf L}_I)^{-1}]_{ij}$, and the Green's function of equation (\ref{expand3}) is written in a matrix form as:
\begin{equation}
i\mathcal{G}_{ij\sigma}^+(z)=\left( \langle e_{i\sigma}| \, \, \, \langle{\Phi_i}| \right) 
i{\sf G} \left(|e_{j\sigma}\rangle \, \, \, |{\Phi_j} \rangle \right)^T,
\label{green2}
\end{equation}
where
${\sf G}=\left(\begin{array}{c c }  {\widehat G}_I & -{\widehat G}_I \\ 0 & {\widehat G}
\end{array}\right)$,  $|\Phi_{j} \rangle={\widehat B}\times i{\widehat G}_{I}|e_{j\sigma}\rangle$,
and the superscript $T$ denotes the transpose.
Using the transformation matrix ${\sf U}=\left(\begin{array}{c c }  I &  {\widehat G}_U \\ 0 & I
\end{array}\right)$, 
where ${\widehat G}_U={\widehat G}_{I}/[{\widehat G}_I-{\widehat G}]$, one can diagonalize ${\sf G}$.
Equation (\ref{green2}) is then rewritten as
\begin{equation}
i\mathcal{G}_{ij\sigma}^+(z)=\left( \langle \widetilde{e}_{i\sigma}| \, \, \,
 \langle{\Phi_i}| \right) i{\sf G}_D \left(
|\widetilde{e}_{j\sigma}\rangle \, \, \, |\Phi_j\rangle \right)^T
\label{green3}
\end{equation}
with the diagonal form ${\sf G}_D=\left(\begin{array}{c c } {\widehat G}_I &  0 \\ 0 &  {\widehat G} \end{array} \right)$,  where
\begin{eqnarray}
|{\tilde e}_{j\sigma}\rangle&=&|e_{j\sigma}\rangle+\frac{{\widehat G}_{I}}{{\widehat G}_{I}-{\widehat G}}|\Phi_j\rangle \nonumber\\
&=&|e_{j\sigma}\rangle+(z{\bf I}/2+i{\bf L}_h)^{-1}(z{\rm\bf I}+i{\rm\bf L}) |\Phi_j\rangle.
\label{newvector}
\end{eqnarray}
Equation (\ref{newvector}) means that the linearly independent components of operators $\widehat{\tilde{e}}_{j\sigma}$ and ${\widehat\Phi}_{j}$ completely span the Liouville space. 

To determine a complete set of basis operators describing $d_\uparrow(t)$, 
we set ${\hat e}_{j\sigma}=d_\uparrow$, and then find  linearly independent operators comprising ${\widehat\Phi}_d$.
For this purpose, we first calculate ${\widehat\Phi}_d=(z{\bf I}/2+i{\bf L}_h) i{\widehat G}_{I}d_\uparrow$.
Since $i{\widehat G}_{I}=(z{\bf I}/2+i{\bf L}_I)^{-1}$, $i{\widehat G}_{I}d_\uparrow$ generates only two linearly independent operators, $d_\uparrow$ and $n_{d\downarrow}d_\uparrow$, using equation (\ref{opexpan}).

Subsequently, we apply $(z{\bf I}/2+i{\bf L}_h)$ to $d_\uparrow$ and $n_{d\downarrow}d_\uparrow$.
The former operation produces $d_\uparrow$ and $c_{k\uparrow}$ with  $k=1, 2, \cdots, \infty$ denoting the quantum states of the metallic reservoir.
Meanwhile, the latter operation $(z{\bf I}/2+i{\bf L}_h) n_{d\downarrow}d_\uparrow$ yields new basis operators $j^-_{d\downarrow}d_\uparrow$, where $j^-_{d\downarrow}=i\sum_k{\tilde{V}}(d^\dagger_\downarrow c_{k\downarrow}-c^\dagger_{k\downarrow}d_\downarrow)$, and $n_{d\downarrow}c_{k\uparrow}$ with  $k=1, 2, \cdots, \infty$ from the operator relation ${\rm\bf L}_{h}n_{d\downarrow}d_\uparrow=[{\rm\bf H}_{h},n_{d\downarrow}]d_\uparrow+n_{d\downarrow}[{\rm\bf H}_{h},d_\uparrow]$.
For simplicity, we consider only one reservoir here and in the following calculations.
Thus, the linearly independent components of ${\widehat \Phi}_d$ are given by
\begin{equation*}
{\widehat \Phi}_d : (d_\uparrow, \, c_{k\uparrow}, \, n_{d\downarrow}d_\uparrow, \, n_{d\downarrow}c_{k\uparrow}, \, j^-_{d\downarrow}d_\uparrow) 
\label{phid}
\end{equation*}
with $\, \, k=1, 2, \cdots, \infty$.

Subsequently, we focus on the operator $\widehat{\tilde{e}}_{j\sigma}$ in equation~(\ref{newvector}). 
First, the operator  ${\bf L}{\widehat \Phi}_d$ consists of five terms
${\bf L}{\widehat \Phi}_d=[{\bf H},d_\uparrow]+[{\bf H},c_{k\uparrow}]+[{\bf H},n_{d\downarrow}d_\uparrow]+[{\bf H},n_{d\downarrow}c_{k\uparrow}]
+[{\bf H},j^-_{d\downarrow}d_\uparrow]$.
We calculate the commutators using an operator identity: $[{\widehat A}{\widehat B},{\widehat C}]={\widehat A}\{{\widehat B},{\widehat C}\}-\{{\widehat A},{\widehat C}\}{\widehat B}$. 
Then, the first and second commutators yield basis operators $d_\uparrow$ and $c_{k\uparrow}$; the third and fourth yield $j^-_{d\downarrow}d_\uparrow$, $n_{d\downarrow}d_\uparrow$, $n_{d\downarrow}c_{k\uparrow}$, and $j^-_{d\downarrow}c_{k\uparrow}$,; and the fifth yields new basis operators 
$j^-_{d\downarrow}n_{d\downarrow}d_\uparrow$, $j^+_{d\downarrow}d_\uparrow$, $j^+_{d\downarrow}n_{d\uparrow}d_\uparrow$, where  
$j^+_{d\downarrow}=\sum_k{\tilde{V}}(d_\downarrow^\dagger c_{k\downarrow}+c^\dagger_{k\downarrow}d_\downarrow)$, and 
$({\bf L}_h j^-_{d\downarrow})d_\uparrow$. 
The last one is discussed below.

Consequently, the linearly independent components of $(z{\bf I}+i{\bf L}){\widehat \Phi}_d$ are classified into two groups: one involving $d_\uparrow$ and the other involving $c_{k\uparrow}$:
\[ 
\{d_\uparrow, n_{d\downarrow}d_\uparrow, j^\mp_{d\downarrow}d_\uparrow, 
j^-_{d\downarrow}n_{d\downarrow}d_\uparrow, j^+_{d\downarrow}n_{d\uparrow}d_\uparrow, ({\bf L}_h j^-_{d\downarrow})d_\uparrow\} 
\]
and
\[
\{c_{k\uparrow}, n_{d\downarrow}c_{k\uparrow}, j^-_{d\downarrow}c_{k\uparrow}\}, \mbox{for} \, k=1, 2, \cdots, \infty.
\]

As a final step to investigate $\widehat{\tilde{e}}_{j\sigma}$, we apply $(z{\bf I}/2+i{\bf L}_h)^{-1}$ to the above operator using equation (\ref{opexpan}), which is an iterative application of ${\bf L}_h$.
One can see that applying ${\rm\bf L}_{h}$ to $d_\uparrow (c_{k\uparrow})$ generates $c_{k\uparrow} (d_\uparrow)$, 
${\rm\bf L}_{h}n_{d\downarrow}$ yields $j^-_{d\downarrow}$, and  ${\rm\bf L}_{h}j^\mp_{d\downarrow}$ yields
\begin{eqnarray*}
{\bf L}_hj^-_{d\downarrow}&=&-i\sum_{\bf l}\sum_{\bf k}({\tilde{V}}^2
c^\dagger_{{\bf l}\downarrow}c_{{\bf
k}\downarrow}+{\tilde{V}}^2c^\dagger_{{\bf
k}\downarrow}c_{{\bf l}\downarrow})\\ &+& 2i\sum_{\bf k}
{\tilde{V}}^2 d^\dagger_{\downarrow}d_\downarrow 
\end{eqnarray*} 
and
\begin{eqnarray*}
{\bf L}_hj^+_{d\downarrow}&=&\sum_{\bf l}\sum_{\bf k}{\tilde{V}}^2
c^\dagger_{{\bf l}\downarrow}c_{{\bf
k}\downarrow}-\sum_{\bf l}\sum_{\bf k}{\tilde{V}}^2
c^\dagger_{{\bf k}\downarrow}c_{{\bf l}\downarrow}.
\end{eqnarray*}
The last two expressions represent a round-trip motion of a down-spin electron between the localized spin and the reservoir.
Therefore, $({\bf L}_h^{n}j^\mp_{d\downarrow})$ signifies the $n+1$ trips of a down-spin electron between the localized spin and the reservoir.

While the full Liouville space provides a detailed description of the dynamics, deriving the Green's function within the entire space is practically impossible.
Therefore, we employ a working Liouville space composed of basis operators, excluding those deemed insignificant.
This simplification is justified by the large-$U$ condition and the condition of unidirectional entangled-state tunneling.

In constructing the working Liouville space, the first requirement eliminates operators $n_{d\downarrow}d_\uparrow$, $j^-_{d\downarrow}n_{d\downarrow}d_\uparrow$, and $j^+_{d\downarrow}n_{d\uparrow}c_{k\uparrow}$ due to strong on-site Coulomb interaction at the localized spin site, while the second requirement eliminates operators  
$j^-_{d\downarrow}j^-_{d\downarrow}d_\uparrow, j^+_{d\downarrow}j^-_{d\uparrow}d_\uparrow$, $({\bf L}_h^{n}j^\mp_{d\downarrow})d_\uparrow$, and $({\bf L}_h^{n}j^\mp_{d\downarrow})c_{k\uparrow}$ with $n\geq 1$.

It is straightforward to extend the above procedure to the case of two reservoirs.
Then, the basis operators are given as: 
\begin{eqnarray*}
{\rm (i)}:&\{&  j^{\pm L}_{d\downarrow}d_{\uparrow}, d_{\uparrow},  j^{\pm R}_{d\downarrow}d_{\uparrow}\}, \\
{\rm (ii)}:&\{& c_{k\uparrow}^{L}, n_{d\downarrow} c_{k\uparrow}^{L},  j^\pm_{d\downarrow}c^R_{k\uparrow},  j^\pm_{d\downarrow}c^L_{k\uparrow}, n_{d\downarrow} c_{k\uparrow}^{R},  c_{k\uparrow}^{R} \}
\end{eqnarray*}
with $k = 0, 1, \cdots, \infty$.

The operators presented above describe all linearly independent ways in which an up-spin at the $d$-level of the central site is annihilated during the entangled-state tunneling under bias, without costing the on-site Coulomb interaction energy, as illustrated in figure~\ref{fig2}. 
In other words, these operators completely span the working Liouville space of $d_\uparrow(t)$ in the systems shown in figure~\ref{fig1}, in which double occupancy is prohibited.

It is worth noting that the basis operators of group (i) describe dynamics in terms of localized spin, while those of group (ii) describe dynamics in terms of electron reservoir.
Thus, group (ii) is used to represent the self-energy.
We adopt the self-energy of the noninteracting Anderson impurity model, which further reduces the working Liouville space by ignoring the operators 
$ j^\pm_{d\downarrow}c^L_{k\uparrow}$ and $j^\pm_{d\downarrow}c^R_{k\uparrow}$.

Thus, the final working Liouville space of this study is spanned by two groups of basis operators:
\begin{eqnarray*}
{\rm I}:&\{& \delta j^{\pm L}_{d\downarrow}d_{\uparrow}, d_{\uparrow},  \delta j^{\pm R}_{d\downarrow}d_{\uparrow}\}, \\
{\rm II}:&\{& c_{k\uparrow}^{L}, \delta n_{d\downarrow} c_{k\uparrow}^{L}, \delta n_{d\downarrow} c_{k\uparrow}^{R}, c_{k\uparrow}^{R} \} \, \mbox{with} \, k = 0, 1, \cdots, \infty.
\end{eqnarray*}
These are the basis operators utilized in section~\ref{sec:formula}.

The basis operators must be orthonormal.
To ensure orthogonality between basis operators, we introduced $\delta$ denoting the fluctuation of an operator about its expectation value.
For normalization, we insert the normalization factor $\langle(\delta n_{d\downarrow})^2\rangle^{1/2}$ or 
$\langle(\delta j^{\pm L/R}_{d\downarrow})^2\rangle^{1/2}$ into the corresponding denominators.

\section{Liouville matrix elements}
\setcounter{section}{3}
To construct the Liouville matrix \(i\mathbf{L}\) using the basis operators given in appendix B, we apply the inner product relationship: $({\hat e}_i,i{\bf L}{\hat e}_j)=-(i{\bf L}{\hat e}_i,{\hat e}_j)$. 
Thus, the matrix element $(i{\bf L})_{ij}$ is thus given by:
\((i\mathbf{L})_{ij} = ({\hat e}_i,i{\bf L}{\hat e}_j) = -(i{\bf L}{\hat e}_i,{\hat e}_j) = -\langle\{i[{\bf H},{\hat e}_i],{\hat e}^\dagger_j\}\rangle\). 
The angular, curly, and square brackets represent the expectation, anticommutator, and commutator, respectively.

Arranging the basis operators in symmetric order, as shown at the top of figure~\ref{figD1} in appendix D, gives the Liouville matrix $i{\bf L}$ composed of nine blocks:
\begin{equation}
i{\bf L}=\left(\begin{array}{c c c} i{\bf L}_{LL} & i{\bf L}_{dL} & {\bf 0} \\
i{\bf L}_{Ld} & i{\bf L}_{dd} & i{\bf L}_{Rd} \\
{\bf 0} & i{\bf L}_{dR} & i{\bf L}_{RR}
\end{array}\right),
\label{eq:Liouville}
\end{equation}
as shown explicitly in figure~\ref{figD1}.

The block $i{\bf L}_{LL}$ ($i{\bf L}_{RR}$) is an $\infty\times\infty$ diagonal block, and the blocks $i{\bf L}_{dL}$ ($i{\bf L}_{dR}$) and $i{\bf L}_{Ld}$ 
($i{\bf L}_{Rd}$) are $5\times\infty$ and $\infty\times 5$ blocks, respectively, while the central block $i{\bf L}_{dd}$ is a $5\times 5$ block.
Moreover, they have the following properties: $i({\bf L}_{dL})=i(-{\bf L}_{Ld}^\dagger)$; $i({\bf L}_{dR})=i(-{\bf L}_{Rd}^\dagger)$; and the block 
$i{\bf L}_{dR}$ ($i{\bf L}_{Rd}$) is point-symmetric with $i{\bf L}_{dL}$ ($i{\bf L}_{Ld}$) about the center of $i{\bf L}$. 
On the other hand, all the blocks surrounding $i{\bf L}_{dd}$ are transformed into the self-energy via the matrix reduction procedure shown in appendix D. 

Subsequently, we present detailed calculations of the matrix elements of $i{\bf L}$ explicitly for the single reservoir Anderson impurity model (SRAIM) for simplicity.
The results for the two-reservoir Anderson impurity model (TRAIM) are provided without detailed procedure. 
The following operator identities are frequently used:
\begin{eqnarray}
	[\hat{A}\hat{B},\hat{C}] &=& \hat{A}\{\hat{B},\hat{C}\}-\{\hat{A},\hat{C}\}\hat{B}, \\
\label{identity0}
%\end{equation}
%\begin{equation}
	\{\hat{A},\hat{B}\hat{C}\} &=& \{\hat{A},\hat{B}\} \hat{C}-\hat{B}[\hat{A},\hat{C}], \quad \mbox{and} \\
\label{identity1}
%\end{equation}
%\begin{eqnarray}
	\{\hat{A}\hat{B},\hat{C}\hat{D}\} &=& \hat{A}[\hat{B},\hat{C}]\hat{D}+\{\hat{A},\hat{C}\}\hat{B}\hat{D}
        +\hat{C}\{\hat{A},\hat{D}\}\hat{B} \nonumber \\
	&-&\hat{C}\hat{A}\{\hat{B},\hat{D}\}.
\label{identity2}
\end{eqnarray}

\subsection{Matrix elements of the block $i{\bf L}_{LL}$}
The block $i{\bf L}_{LL}$ ($i{\bf L}_{RR}$) is composed of two infinite-dimensional diagonal blocks with elements $i\epsilon_k$
that are constructed by the basis operators $c_{k\uparrow}^{\nu}$
and $\delta n_{d\downarrow}c_{k\uparrow}^{\nu}$ with $k=0, 1, \ldots, \infty$ describing the
$\nu$ reservoir.
We skip the calculation procedure because it is simple.

\subsection{Matrix elements of the block $i{\bf L}_{dd}$}

We set matrix elements of the central block $i{\bf L}_{dd}$ as follows:
\begin{eqnarray}
i{\bf L}_{dd}=\left( \begin{array}{c c c c c} D_1 & \gamma^L &
-U^L_{j^-} & \gamma^{LR}_S & \gamma^{LR}_A \\ -\gamma^L & D_2
& -U^L_{j^+} & \gamma^{LR}_A & \gamma^{LR}_S \\
U_{j^-}^{L*} &  U_{j^+}^{L*} & D_3 &  U^{R*}_{j^+} &
U^{R*}_{j^-} \\  -\gamma^{LR}_S & -\gamma^{LR}_A & -U_{j^+}^R  &
D_4 & -\gamma^R \\
 -\gamma^{LR}_A &  -\gamma^{LR}_S &  -U_{j^-}^R  & \gamma^R & D_5
\end{array} \right)
\label{center}
\end{eqnarray}
We derive the expression of each matrix element through the calculations for the SRAIM.

\subsubsection{Diagonal elements.}

The diagonal elements are given by $D_{1,2}=\langle\{i{\bf L}(\delta j^\mp_{d\downarrow}d_{\uparrow}), \delta j^\mp_{d\downarrow}d_{\uparrow}^\dagger\}\rangle$ and $D_3=-\langle\{i{\bf L}d_{\uparrow}, d_{\uparrow}^\dagger\}\rangle$ for the SRAIM.
Using the commutator $i[{\bf H},d_{\uparrow}]=-i\sum_{\bf
k}\tilde{V} c_{{\bf k}\uparrow}-i\epsilon_d d_{\uparrow}-iU
d_{\uparrow} n_{d\downarrow}$ gives the diagonal elements as follows:
\begin{eqnarray*}
	D_3: \quad
	&-&\langle\{i[{\bf H},d_{\uparrow}], d^\dagger_{\uparrow}\}\rangle= i\sum_{\bf k}\tilde{V} \langle\{ c_{{\bf k}\uparrow},
		d^\dagger_{\uparrow}\}\rangle\\
       &+&i\epsilon_d\langle\{ d_{\uparrow},d^\dagger_{\uparrow}\}\rangle+iU\langle\{d_{\uparrow} n_{d\downarrow},d^\dagger_{\uparrow}\}\rangle \\
       &=& i\epsilon_d+iU\langle\{d_{\uparrow}n_{d\downarrow},d^\dagger_{\uparrow}\}\rangle \\
       &=&i\epsilon_d+iU\langle\{d_{\uparrow},d^\dagger_{\uparrow}\} n_{d\downarrow}\rangle=i\epsilon_d+iU\langle n_{d\downarrow}\rangle,
\end{eqnarray*}
and
\begin{eqnarray*}
	D_{1,2}: \quad  
       &-&\langle\{i[{\bf H},d_{\uparrow}\delta j^\mp_{d\downarrow}],(d_{\uparrow}\delta j^\mp_{d\downarrow})^\dagger\}\rangle\\
	&=&-\langle\{i[{\bf H},d_{\uparrow}]\delta j^\mp_{d\downarrow}, \delta j^\mp_{d\downarrow}d_{\uparrow}^\dagger\}\rangle\\
        &-&\langle\{d_{\uparrow}i[{\bf H},\delta j^\mp_{d\downarrow}], \delta j^\mp_{d\downarrow}d_{\uparrow}^\dagger\}\rangle.
\end{eqnarray*}
The first term of $D_{1,2}$ is rewritten as
\begin{eqnarray*}
	&-&\langle\{i[{\bf H},d_{\uparrow}]\delta j^\mp_{d\downarrow},
		\delta j^\mp_{d\downarrow}d_{\uparrow}^\dagger\}\rangle\\
	&=&\langle\{(i\sum_{\bf k}\tilde{V}c_{{\bf k}\uparrow}
		+i\epsilon_d d_{\uparrow}
		+iU d_{\uparrow} n_{d\downarrow}) \delta j^\mp_{d\downarrow},
		(d_{\uparrow}\delta j^\mp_{d\downarrow})^\dagger\}\rangle \\
	&=&i\epsilon_d\langle\{d_{\uparrow}\delta j^\mp_{d\downarrow},
		\delta j^\mp_{d\downarrow}d_{\uparrow}^\dagger\}\rangle
		+iU\langle\{d_{\uparrow}n_{d\downarrow}\delta
		j^\mp_{d\downarrow},\delta j^\mp_{d\downarrow}d_{\uparrow}^\dagger\}\rangle.
\end{eqnarray*}
Applying the decoupling approximation, $n_{d\downarrow}\delta
j^\mp_{d\downarrow}=\langle n_{d\downarrow}\rangle\delta
j^\mp_{d\downarrow}$, to the $U$-term above gives rise to a form
with squared norm:
\begin{equation*}
	-\langle\{i[{\bf H},d_{\uparrow}]\delta j^\mp_{d\downarrow}, \delta
	j^\mp_{d\downarrow}d_{\uparrow}^\dagger\}\rangle \\
	=[i\epsilon_d+iU\langle n_{d\downarrow}\rangle]\times
	||d_{\uparrow}\delta j^-_{d\downarrow}||^2.
\end{equation*}
Meanwhile, the second term of $D_{1,2}$ vanishes using the relation $[{\bf H},j^\mp_{d\downarrow}]\propto j^\pm_{d\downarrow}$ and the
orthogonality condition of the basis operators.
Thus, the diagonal elements of the block $i{\bf L}_{dd}$ for the SRAIM are $i\epsilon_d+iU\langle n_{d\downarrow}\rangle$.

\subsubsection{Matrix elements $U_{j^\mp}^{\nu}$.}

The elements $U_{j^\mp}$ for the SRAIM come from the inner product $(d_{\uparrow}, i{\bf L}\delta j^\mp_{d\downarrow}d_{\uparrow}^\dagger)=-(i{\bf L}d_{\uparrow}, \delta j^\mp_{d\downarrow}d_{\uparrow}^\dagger)$, which is given by
\begin{eqnarray*}
&-&(i{\bf L}d_{\uparrow}, \delta j^\mp_{d\downarrow}d_{\uparrow}^\dagger)=-\langle\{ i[{\bf H},d_{\uparrow}], \delta j^\mp_{d\downarrow}
		d_{\uparrow}^\dagger\}\rangle \\
        &=& i\epsilon_d\langle\delta j^\mp_{d\downarrow}\rangle\langle\{d_{\uparrow},d_{\uparrow}^\dagger\}\rangle
	+iU\langle\{d_{\uparrow}n_{d\downarrow},\delta j^\mp_{d\downarrow} d_{\uparrow}^\dagger\}\rangle\\
	&+&i\sum_{\bf k}\tilde{V}\langle\delta j^\mp_{d\downarrow} \rangle
		\langle\{c_{{\bf k}\uparrow},d_{\uparrow}^\dagger\}\rangle \\
	&=& iU\langle\{d_{\uparrow}n_{d\downarrow},\delta j^\mp_{d\downarrow}
		d_{\uparrow}^\dagger\}\rangle.
\end{eqnarray*}
The operator identity of equation~(\ref{identity2}) gives $\langle\{d_{\uparrow}n_{d\downarrow},\delta j^\mp_{d\downarrow}
d_{\uparrow}^\dagger\}\rangle$ as follows:
\begin{eqnarray*} 
&& \langle\{ d_{\uparrow}\delta n_{d\downarrow},\delta
	 j^\mp_{d\downarrow}d_{\uparrow}^\dagger\}\rangle
         =\langle d_{\uparrow}d_{\uparrow}^\dagger[n_{d\downarrow},
	 j^\mp_{d\downarrow}]\rangle+\langle\delta
  	 j^\mp_{d\downarrow}\delta n_{d\downarrow}\rangle \\
      	&=&\langle(1-n_{d\uparrow})[n_{d\downarrow},j^\mp_{d\downarrow}]
		+\delta j^\mp_{d\downarrow}\delta n_{d\downarrow}\rangle \\
	&=& \langle(\frac{1}{2}-n_{d\uparrow})[n_{d\downarrow},
		j^\mp_{d\downarrow}]+\frac{1}{2}[\delta n_{d\downarrow},
		\delta j^\mp_{d\downarrow}]+\delta j^\mp_{d\downarrow} 
		\delta n_{d\downarrow}\rangle \\
	&=& \frac{1}{2}\langle(1-2n_{d\uparrow})[n_{d\downarrow},
		j^\mp_{d\downarrow}]+\{\delta n_{d\downarrow},
		\delta j^\mp_{d\downarrow}\}\rangle \\
	&=&\frac{1}{2}\{\langle(1-2n_{d\uparrow})[n_{d\downarrow},
		j^\mp_{d\downarrow}]\rangle+(1-2\langle n_{d\downarrow}\rangle)
		\langle j^\mp_{d\downarrow}\rangle\}.
\end{eqnarray*}
Hence, the matrix elements $U_{j^\mp}^\nu$  for the TRAIM is given by
\begin{eqnarray}
\frac{U_{j^\mp}^{\nu}}{U}&=&\frac{\langle
i(1-2n_{d\uparrow})[n_{d\downarrow},j^{\mp \nu}_{d\downarrow}]\rangle+i(1-2\langle n_{d\downarrow}\rangle)\langle j^{\mp
\nu}_{d\downarrow}\rangle}{2\langle(\delta j^{\mp
\nu}_{d\downarrow})^2\rangle^{1/2}}. \nonumber \\
\label{eq:Ualphaj}
\end{eqnarray}
We added the denominator explicitly.
Note that the second term, the imaginary part, reflects the deviation from half-filling.

\subsubsection{Matrix elements $\gamma^{L,R}$, $\gamma^{LR}_S$, and $\gamma^{LR}_A$.}

Matrix element $\gamma$ for the SRAIM is given by
\begin{eqnarray}
	&-&\langle\{ i{\bf L}(d_{\uparrow}\delta j^\mp_{d\downarrow}),
	\delta j^\pm_{d\downarrow}d_{\uparrow}^\dagger\}\rangle
	=-\langle\{i[{\bf H},d_{\uparrow}]\delta j^\mp_{d\downarrow},
	\delta j^\pm_{d\downarrow}d_{\uparrow}^\dagger\}\rangle \nonumber\\
	&-&\langle\{d_{\uparrow}i[{\bf H}, j^\mp_{d\downarrow}],
	\delta j^\pm_{d\downarrow}d_{\uparrow}^\dagger\}\rangle.
\label{eq:gamma11}
\end{eqnarray}
The second term vanishes because the inner product contains a single $\delta j^\pm_{d\downarrow}$, while the first term is rewritten as
\begin{eqnarray}
	&-&\langle\{i[{\bf H},d_{\uparrow}]\delta j^\mp_{d\downarrow},
	\delta j^\pm_{d\downarrow}d_{\uparrow}^\dagger\}\rangle\nonumber\\
	&=&i\epsilon_d\langle\{d_{\uparrow}\delta j^\mp_{d\downarrow},
	\delta j^\pm_{d\downarrow}d_{\uparrow}^\dagger\}\rangle
	+iU\langle\{d_{\uparrow}n_{d\downarrow}\delta j^\mp_{d\downarrow},
	\delta j^\pm_{d\downarrow}d_{\uparrow}^\dagger\}\rangle \nonumber \nonumber\\
	&+& i\sum_{\bf k}\tilde{V}\langle\{c_{k\uparrow}\delta j^\mp_{d\downarrow},
	\delta j^\pm_{d\downarrow}d_{\uparrow}^\dagger\}\rangle.
  \label{eq:gamma22}
\end{eqnarray}
The first and second terms of equation~(\ref{eq:gamma22}) vanish by applying the
orthogonality condition of the basis operators,
	$\langle \{d_{\uparrow}\delta j^\pm_{d\downarrow},
	\delta j^\mp_{d\downarrow}d_{\uparrow}^\dagger\}\rangle=0 $,
and decoupling approximation $\langle n_{d\downarrow}\rangle$ to the second term.
Meanwhile, the third term of equation~(\ref{eq:gamma22}) is rewritten as
\begin{eqnarray*}
	&&i\sum_{\bf k}\tilde{V}
		\langle\{ c_{k\uparrow}\delta j^\mp_{d\downarrow},
		\delta j^\pm_{d\downarrow}d_{\uparrow}^\dagger\}\rangle\\
	&=&i\sum_{\bf k}\tilde{V}
		\langle\{ \delta j^\mp_{d\downarrow}\delta j^\pm_{d\downarrow}
		c_{k\uparrow}d_{\uparrow}^\dagger
		-\delta j^\pm_{d\downarrow}\delta j^\mp_{d\downarrow}
		c_{k\uparrow}d_{\uparrow}^\dagger\}\rangle \\
	&=&i\sum_{\bf k}\tilde{V}\langle[\delta j^\mp_{d\downarrow},
		\delta j^\pm_{d\downarrow}]
		c_{k\uparrow}d_{\uparrow}^\dagger \rangle=i\sum_{\bf k}\tilde{V}\langle[j^\mp_{d\downarrow},
		j^\pm_{d\downarrow}]c_{k\uparrow}d_{\uparrow}^\dagger \rangle.
\end{eqnarray*}
Finally, the matrix element $\gamma$ for the SRAIM is given by
\begin{equation*}
	\gamma=\frac{-\langle\{i{\bf L}(d_{\uparrow}\delta j^{\mp}_{d\downarrow}),
		\delta j^{\pm}_{d\downarrow}d_{\uparrow}^\dagger\}\rangle}
	{\sqrt{\langle( \delta j^{\mp}_{d\downarrow})^2\rangle}
	\sqrt{\langle( \delta j^{\pm}_{d\downarrow})^2\rangle}}
	=\frac{i\sum_{\bf k}\tilde{V}\langle[j^{\mp}_{d\downarrow},
	j^{\pm}_{d\downarrow}]c_{k\uparrow}d_{\uparrow}^\dagger \rangle}
	{\sqrt{\langle( \delta j^{\mp}_{d\downarrow})^2\rangle}
	\sqrt{\langle( \delta j^{\pm}_{d\downarrow})^2\rangle}}.
\end{equation*}
For TRAIM, applying substitutions such as $\tilde{V}c_{k\uparrow}\rightarrow (\tilde{V}^Lc_{k\uparrow}^L+\tilde{V}^Rc_{k\uparrow}^R)$ and $j^{\mp}_{d\downarrow}\rightarrow j^{\mp \nu}_{d\downarrow}$ for $\gamma^{\nu}$ (where $\nu \in L, R$) yields the correct expressions for $\gamma^L$ and $\gamma^R$.
In contrast, to obtain the expressions for $\gamma^{LR}_{S,A}$ presented in equations (\ref{gamma-S}) and (\ref{gamma-A}) of the main text, additional considerations—such as the combinations of the superscripts ($L, R$) and ($+,-$) for $j$—are required.

\subsection{Matrix elements of the block $i{\bf L}_{dL}$}
The block $i{\bf L}_{dL}$ is written as
\begin{eqnarray}
i{\bf L}_{dL}=\left[ \begin{array}{c c c c c}  {\bf 0} & {\bf 0} & i{\bf C}^L_{{\bf k}d} & {\bf 0} &  {\bf 0} \\
	i{\bf C}_{{\bf k}j^-}^{\, LL} & i{\bf C}_{{\bf k}j^+}^{\, LL} &  {\bf 0}
 & i{\bf C}_{{\bf k}j^+}^{\, LR} & i{\bf C}_{{\bf k}j^-}^{\, LR}
\end{array} \right], \nonumber \\
\label{eq:Lml}
\end{eqnarray}
where $i{\bf C}^L_{{\bf k}d}$, $i{\bf C}_{{\bf k}j^\mp}^{LL}$, and
$i{\bf C}_{{\bf k}j^\mp}^{LR}$ are infinite-dimensional column vectors, as shown in figure~\ref{figD1}.
The elements of $i{\bf C}^L_{{\bf k}d}$ and $i{\bf C}_{{\bf k}j^\mp}^{LL}$ are given by 
$-\langle\{i{\bf L}c_{k\uparrow}, d_{\uparrow}^\dagger\}\rangle$ and $-\langle\{i{\bf L}(c_{k\uparrow}\delta n_{d\downarrow}), \delta
j^\mp_{d\downarrow}d_{\uparrow}^\dagger\}\rangle$, respectively. Note that calculations are done for the SRAIM.
Thus, they are given as follows:
\begin{eqnarray}
i{\bf C}_{{\bf k}d}&=&-\langle\{i{\bf L}c_{k\uparrow}, d_{\uparrow}^\dagger\}\rangle=-\langle\{i[{\bf H},c_{k\uparrow}], d_{\uparrow}^\dagger\}\rangle \nonumber \\
&=&i\epsilon_k\langle\{c_{k\uparrow}, d_{\uparrow}^\dagger\}\rangle+i\tilde{V}\langle\{d_{\uparrow}, d_{\uparrow}^\dagger\}\rangle
=i\tilde{V}
\label{eq:tildeV}
\end{eqnarray}
 and  
\begin{eqnarray}
	i{\bf C}_{{\bf k}j^\mp}&=&-\langle\{i{\bf L}(c_{k\uparrow}\delta n_{d\downarrow}), 
\delta j^\mp_{d\downarrow}d_{\uparrow}^\dagger\}\rangle \nonumber\\
&=&-\langle\{i[{\bf H},c_{k\uparrow}\delta n_{d\downarrow}],
		\delta j^\mp_{d\downarrow}d_{\uparrow}^\dagger\}\rangle \nonumber \\
	&=&-\langle\{i[{\bf H},c_{k\uparrow}]\delta
		n_{d\downarrow}+c_{k\uparrow}i[{\bf H},n_{d\downarrow}],
		\delta j^\mp_{d\downarrow}d_{\uparrow}^\dagger\}\rangle \nonumber \\
	&=&i\epsilon_k\langle\{c_{k\uparrow}\delta n_{d\downarrow},
		\delta j^\mp_{d\downarrow}d_{\uparrow}^\dagger\}\rangle
		+i\tilde{V}\langle\{d_{\uparrow}\delta n_{d\downarrow},
		\delta j^\mp_{d\downarrow}d_{\uparrow}^\dagger\}\rangle \nonumber \\
	&-&\langle\{c_{k\uparrow}j^-_{d\downarrow}, \delta j^\mp_{d\downarrow}d_{\uparrow}^\dagger\}\rangle.
\label{eq:cll-k}
\end{eqnarray}
The first term $i\epsilon_k\langle\{c_{k\uparrow}\delta
n_{d\downarrow},\delta j^\mp_{d\downarrow}d_{\uparrow}^\dagger\}\rangle$ in equation~(\ref{eq:cll-k})
vanishes due to the orthogonality condition of  basis operators.
The third term is rewritten as a combination of terms containing commutators and anticommutators using equation~(\ref{identity2}).
One can easily find that all the commutators and anticommutators vanish.

In contrast, the second term must be rigorously treated because it is non-vanishing. 
Using equation~(\ref{identity2}), the second term $i\tilde{V}\langle\{ d_{\uparrow}\delta n_{d\downarrow},\delta  j^\mp_{d\downarrow}d_{\uparrow}^\dagger\}\rangle$ is rewritten as

$\langle\{ d_{\uparrow}\delta n_{d\downarrow},\delta  j^\mp_{d\downarrow}d_{\uparrow}^\dagger\}\rangle
         =\langle d_{\uparrow}d_{\uparrow}^\dagger[n_{d\downarrow},  j^\mp_{d\downarrow}]\rangle+\langle\delta
  	 j^\mp_{d\downarrow}\delta n_{d\downarrow}\rangle\\
        =\langle(1-n_{d\uparrow})[n_{d\downarrow},j^\mp_{d\downarrow}]	+\delta j^\mp_{d\downarrow}\delta n_{d\downarrow}\rangle\\
        = \langle(\frac{1}{2}-n_{d\uparrow})[n_{d\downarrow},	j^\mp_{d\downarrow}]+\frac{1}{2}[\delta n_{d\downarrow}, \delta j^\mp_{d\downarrow}]+\delta j^\mp_{d\downarrow}\delta n_{d\downarrow}\rangle \\
       = \frac{1}{2}\langle(1-2n_{d\uparrow})[n_{d\downarrow}, j^\mp_{d\downarrow}]+\{\delta n_{d\downarrow}, \delta j^\mp_{d\downarrow}\}\rangle\\
=\frac{1}{2}\{\langle(1-2n_{d\uparrow})[n_{d\downarrow}, j^\mp_{d\downarrow}]\rangle+
              (1-2\langle n_{d\downarrow}\rangle) \langle j^\mp_{d\downarrow}\rangle\}.$

Thus, the matrix elements of $i{\bf C}_{{\bf k}j^\mp}$ in the block $i{\bf L}_{dL}$ for the SRAIM are written as:
\begin{eqnarray}
	&i{\bf C}_{{\bf k}j^\mp}&=-\langle\{i{\bf L}(c^L_{k\uparrow}\delta n_{d\downarrow}),
	\delta j^\mp_{d\downarrow}d_{\uparrow}^\dagger\} \rangle \nonumber \\
	&=&\frac{\tilde{V}^L}{2}\frac{\langle i(1-2n_{d\uparrow})
	[n_{d\downarrow},j^\mp_{d\downarrow}]\rangle
	+i(1-2\langle n_{d\downarrow}\rangle)\langle
	j^\mp_{d\downarrow}\rangle}{\langle(\delta n_{d\downarrow})^2\rangle^{1/2}
	\langle(\delta j^\mp_{d\downarrow})^2\rangle^{1/2}} \nonumber \\
&\equiv& \tilde{V}^L\xi_{\mp 1},
\label{eq:xi-1}
\end{eqnarray}
where the second term of the numerator is the imaginary part of $\tilde{V}^L\xi_{\mp 1}$, which vanishes at half-filling, as is the case in this study.
Here, we explicitly include the reservoir index $L$ in $\tilde{V}$ to indicate that $\tilde{V}^L$ originates from the operator $c^L_{k\uparrow}$ rather than $j^\mp_{d\downarrow}$.

The third term of Eq. (\ref{eq:cll-k}), $-\langle\{c_{k\uparrow}j^-_{d\downarrow},\delta j^\mp_{d\downarrow}d_{\uparrow}^\dagger\}\rangle$, becomes $-\langle\{c_{k\uparrow}^L(j^{-L}_{d\downarrow}+j^{-R}_{d\downarrow}),\delta j^{\mp L}_{d\downarrow}d_{\uparrow}^\dagger\}\rangle$ for TRAIM.
It is noteworthy that the second part, $-\langle\{c_{k\uparrow}^Lj^{-R}_{d\downarrow}, \delta j^{\mp L}_{d\downarrow}d_{\uparrow}^\dagger\}\rangle$, reflects nonvanishing left-right entanglement, which is expressed as $\tilde{V}^\nu\xi^{\nu\nu'}_{\mp 2}$ in what follows. 
Here, too, the reservoir index of $\tilde{V}^\nu$ originates from $c_{k\uparrow}$.
Consequently,  for the TRAIM,
\begin{equation}
i{\bf C}_{{\bf k}j^\mp}^{\nu\nu'}=\tilde{V}^\nu(\xi^{\nu\nu'}_{\mp 1} + \xi^{\nu\nu'}_{\mp 2})\equiv \tilde{V}^\nu\xi^{\nu\nu'}_{\mp}.
\label{eq:xi-12}
\end{equation} 

The first term, $\tilde{V}^\nu\xi^{\nu\nu'}_{\mp 1}$, is given in equation~(\ref{eq:xi-1}), and the second term, $\tilde{V}^\nu\xi^{\nu\nu'}_{\mp 2}$, is written as:
\begin{eqnarray}
\tilde{V}^L\xi_{-2}^{LL} &=& \frac{2\langle\{c_{k\uparrow}^L(j^{-L}_{d\downarrow}+j^{-R}_{d\downarrow}),
	\delta j^{-L}_{d\downarrow} d_{\uparrow}^{\dagger}\}\rangle^{\rm sn}}{\langle(\delta j^{-L}_{d\downarrow})^2\rangle^{1/2}} \nonumber \\
&=&\frac{2\langle c_{k\uparrow}^L d_{\uparrow}^{\dagger} 
		[j^{-R}_{d\downarrow}, j^{-L}_{d\downarrow}]\rangle^{\rm sn}}{\langle(\delta j^{-L}_{d\downarrow})^2\rangle^{1/2}} \, \, \, \mbox{from} \, \, \,  i{\bf C}_{{\bf k}j^-}^{LL}, 
\label{eq:xi-2-1}
\end{eqnarray}
\begin{eqnarray}
\tilde{V}^L\xi_{-2}^{LR}&=& \frac{2\langle\{c_{k\uparrow}^L(j^{-L}_{d\downarrow}+j^{-R}_{d\downarrow}),
	\delta j^{-R}_{d\downarrow} d_{\uparrow}^{\dagger}\}\rangle^{\rm sn}}{\langle(\delta j^{-L}_{d\downarrow})^2\rangle^{1/2}} \nonumber \\
&=&\frac{2\langle c_{k\uparrow}^Ld_{\uparrow}^{\dagger}
	[j^{-L}_{d\downarrow}, j^{-R}_{d\downarrow}]\rangle^{\rm sn}}{\langle(\delta j^{-R}_{d\downarrow})^2\rangle^{1/2}}   \, \, \, \mbox{from} \, \, \, i{\bf C}_{{\bf k}j^-}^{LR}, 
\label{eq:xi-2-2}
\end{eqnarray}
\begin{eqnarray}
\tilde{V}^L\xi_{+2}^{LL} &=& \frac{2\langle\{c_{k\uparrow}^L(j^{-L}_{d\downarrow}+j^{-R}_{d\downarrow}),
	\delta j^{+L}_{d\downarrow} d_{\uparrow}^{\dagger}\}\rangle^{\rm sn}}{\langle(\delta j^{+L}_{d\downarrow})^2\rangle^{1/2}} \nonumber \\
&=& \frac{2\langle c_{k\uparrow}^L d_{\uparrow}^{\dagger} 
		[j^{-L}_{d\downarrow},j^{+L}_{d\downarrow}]\rangle^{\rm sn}}{\langle(\delta j^{+L}_{d\downarrow})^2\rangle^{1/2}} \nonumber \\
		&+&\frac{2\langle c_{k\uparrow}^L d_{\uparrow}^{\dagger} 
		[j^{-R}_{d\downarrow},j^{+L}_{d\downarrow}]\rangle^{\rm sn}}{\langle(\delta j^{+L}_{d\downarrow})^2\rangle^{1/2}} 
 \mbox{from} \, \, \, i{\bf C}_{{\bf k}j^+}^{LL},  
\label{eq:xi-2-3} 
\end{eqnarray}
\begin{eqnarray}
\tilde{V}^L\xi_{+2}^{LR}&=& \frac{2\langle\{c_{k\uparrow}^L(j^{-L}_{d\downarrow}+j^{-R}_{d\downarrow}),
	\delta j^{+R}_{d\downarrow} d_{\uparrow}^{\dagger}\}\rangle^{\rm sn}}{\langle(\delta j^{+L}_{d\downarrow})^2\rangle^{1/2}} \nonumber \\
&=&\frac{2\langle c_{k\uparrow}^L d_{\uparrow}^{\dagger} 
		[j^{-R}_{d\downarrow}, j^{+R}_{d\downarrow}] \rangle^{\rm sn}}{\langle(\delta j^{+R}_{d\downarrow})^2\rangle^{1/2}} \nonumber \\
		&+&\frac{2\langle c_{k\uparrow}^L d_{\uparrow}^{\dagger} 
		[j^{-L}_{d\downarrow}, j^{+R}_{d\downarrow}]\rangle^{\rm sn}}{\langle(\delta j^{+R}_{d\downarrow})^2\rangle^{1/2}}  
 \mbox{from} \, \, \, i{\bf C}_{{\bf k}j^+}^{LR}.
\label{eq:xi-2-4}
\end{eqnarray}

Note that \(\tilde{V}^\nu\xi^{\nu\nu'}_{\mp 2}\) are quantities obtained at steady-state non-equilibrium.
Therefore, the superscript 'sn' is used.
The dynamics occurring in \(\tilde{V}^\nu\xi^{\nu\nu'}_{\mp 2}\) are substantially the same as those of the $\gamma$'s in equations~(\ref{gamma-L}), (\ref{gamma-S}), and (\ref{gamma-A}), whose entangled-state tunneling dynamics are partly illustrated in figure~\ref{fig2}.

Notably, the above expressions yield the inequality:
\begin{equation}
|\xi_{-2}^{LL}|=|\xi_{-2}^{LR}|<|\xi_{+2}^{LL}|=|\xi_{+2}^{LR}|,
\label{eq:inequality}
\end{equation}
which is used in determining the inequalities among the self-energy coefficients, $\eta_{pq}$, in appendix F.

\section{Matrix reduction}
\setcounter{section}{4} 

The on-site retarded Green's function given in equations (\ref{eq:GR04}) in appendix A and (\ref{green0}) in the main text,
$$
i\mathcal{G}_{dd}^{+}(z) = \left[\frac{1}{z \mathbf{I} + i \mathbf{L}}\right]_{dd},
$$
can be expressed in a matrix form as

${\rm\bf M}{\bf A}_c={\bf D}_c \, \, \mbox{or} \, \, {\bf A}_c= {\rm\bf M}^{-1}{\bf D}_c,$

\noindent where ${\bf A}_c$ and ${\bf D}_c$ are column vectors defined as ${\bf A}_c=$

\noindent $(A_{\infty L}(z),\cdots\!,A_{1L}(z),i\mathcal{G}_{dd}^{+}(z), A_{1R}(z),\cdots\!,A_{\infty R}(z))^T$ 

\noindent and ${\bf D}_c= (0,\cdots,0,1, 0,\cdots,0)^T$.

\begin{figure}[t] 
\centering
\includegraphics[width=3.0 in]{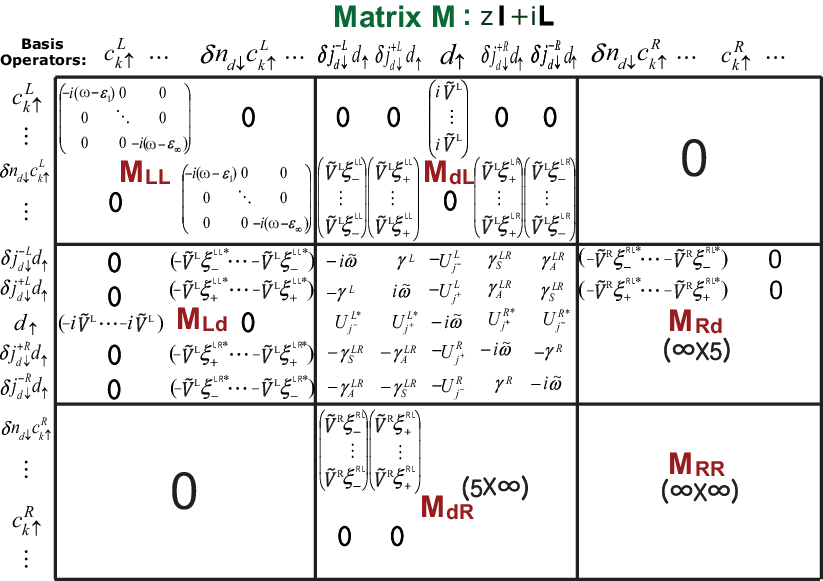}
\caption{
Block structure of the matrix ${\rm\bf M}$ shown in figure~\ref{fig6}. The nine submatrices are indicated by the thick black lines.
}
\label{figD1}
\end{figure}

As shown in figure~\ref{figD1}, the matrix \({\bf M}\) consists of nine blocks, such as \({\bf M}_{LL}\) (\(\infty \times \infty\)), \({\bf M}_{Ld}\) (\(5 \times \infty\)), \({\bf M}_{dL}\) (\(\infty \times 5\)), and \({\bf M}_{dd}\) (\(5 \times 5\)). 
The matrix equation is thus expressed as:  

\begin{equation}
\left(\begin{array}{ccc} {\bf M}_{LL} & {\bf M}_{dL} & {\bf 0}  \\ 
{\bf M}_{Ld} & {\bf M}_{dd} & {\bf M}_{Rd}  \\
{\bf 0} &  {\bf M}_{dR} & {\bf M}_{RR}
\end{array} \right)
\left(\begin{array}{c} {\bf C}^L \\ {\bf C}^d \\ {\bf C}^R
\end{array} \right)
=\left(\begin{array}{c}{\bf 0} \\ {\bf I}_5 \\ {\bf 0}
\end{array} \right)
\label{eq:M-matrix}
\end{equation}
where \({\bf C}^L = (A_{\infty L}(z), \cdots, A_{3L}(z))^T\), \({\bf C}^d = (A_{2L}(z), A_{1L}(z), i{\mathcal G}^+_{dd}(z), A_{1R}(z), A_{2R}(z))^T\), \({\bf C}^R = (A_{3R}(z), \cdots, A_{\infty R}(z))^T\), \({\bf 0}\) is an infinite-dimensional zero vector, and \({\bf I}_{5} = (0 \, 0 \, 1 \, 0 \, 0)^T\).  
Eliminating ${\bf C}^L$ and ${\bf C}^R$ from the three coupled equations obtained from equation~(\ref{eq:M-matrix}) gives the matrix equation to obtain the reduced matrix:  
\begin{equation}
({\bf M}_{dd}-{\bf M}_{Ld}{\bf M}_{LL}^{-1}{\bf M}_{dL} -
	{\bf M}_{Rd}{\bf M}_{RR}^{-1}{\bf M}_{dR}){\bf C}_d={\bf I}_5.
\label{eq:M-matrix11}
\end{equation}

Since \({\bf M}_{LL}\) and \({\bf M}_{RR}\) are diagonal matrices with elements $(-i\omega+ i\epsilon_k)$, where $k = 0, 1, \cdots, \infty$, 
their inverses can be evaluated straightforwardly.
Consequently, the latter two terms in parentheses in equation~(\ref{eq:M-matrix11}) can be expressed as a \(5 \times 5\) matrix with self-energy elements \(i{\bf \Sigma}_{pq} = \eta_{pq}[i{\bf \Sigma}_0^L(\omega) + i{\bf \Sigma}_0^R(\omega)]\),
where $i{\bf \Sigma}_0^\nu(\omega)=i\Sigma_{\bf k}(\tilde{V}^{\nu})^2/(\omega-\epsilon_{\bf k})=\pi(\tilde{V}^\nu)^2\rho^{\nu}_0(\omega)$, and $\rho^{\nu}_0(\omega)$ denotes the density of states of the $\nu$ reservoir.
The coefficients \(\eta_{pq}\) are detailed in appendix F below.   

Thus, the equivalent \(5 \times 5\) matrix equation becomes:  
\( {\bf M}_r {\bf C}_d = {\bf I}_5,\)  
where \({\bf M}_r = {\bf M}_{dd} + i{\bf \Sigma}\) and \({\bf C}^d = (A_{2L}(z), A_{1L}(z), i{\mathcal G}^+_{dd}(z), A_{1R}(z), A_{2R}(z))^T\).
This matrix equation writes the on-site retarded Green’s function as:  
\begin{equation}
	i{\mathcal G}^{+}_{dd}(\omega)=({\bf M}_{r}^{-1})_{33}\equiv({\bf M}_{r}^{-1})_{dd},
\label{eq:M-matrix2}
\end{equation}  
and the local density of states for an $\uparrow$-spin at $d$, \(\rho_{d\uparrow}(\omega)\), is given by:  
\begin{equation}
	\rho_{d\uparrow}(\omega)=-\frac{1}{\pi}{\rm Im}{\mathcal G}^{+}_{dd\uparrow}(\omega)=\frac{1}{\pi}{\rm Re}({\bf M}_{r}^{-1})_{33}.
\label{eq:M-matrix3}
\end{equation}
This expression gives the differential conductance in equation~(\ref{eq:new-dIdV}) in the main text.

%%%%%%%%%%%%%%%%%%%%%%%%%%%%%%%%%%%%%%%%%%%%%%%%%%%%%%%%%%%%%%%%%%%%%%%%%%%%%%%%%%%%%%%%
%%%%%%%%%%%%%%%%%%%%%%%%%%%%%%%%%%%%%%%%%%%%%%%%%%%%%%%%%%%%%%%%%%%%%%%%%%%%%%%%%%%%%%%%

\section{Entangled-state tunneling}
\setcounter{section}{5} 
In \(\gamma^L\) of equation~(\ref{gamma-L}), for example, the commutator is given by
\begin{eqnarray}
[j^{-L}_{d\downarrow},j^{+L}_{d\downarrow}]=2i\sum_{k\in
L}({\tilde{V}}^L)^2(c^{L\dagger}_{k\downarrow}d_\downarrow d^\dagger_\downarrow c^L_{k\downarrow}
-d^\dagger_\downarrow c^L_{k\downarrow}c^{L\dagger}_{k\downarrow}d_\downarrow). \nonumber\\
\label{LL}
\end{eqnarray}
Unidirectional entangled-state tunneling from the left to the right reservoir is established by selecting four $\downarrow$-spin operators,
$c^{L\dagger}_{k\downarrow}d_\downarrow d^\dagger_\downarrow c^{L}_{k\downarrow}$, from the commutator (\ref{LL}),
two $\uparrow$-spin operators $c_{k\uparrow}^L d^\dagger_\uparrow$ from $(\tilde{V}^L c_{k\uparrow}^L+\tilde{V}^R c^R_{k'\uparrow})d^\dagger_{\uparrow}$ in the numerator of equation~(\ref{gamma-L}), and  two $\uparrow$-spin operators
$c_{k'\uparrow}^{R\dagger} d_\uparrow$ from the definition of the steady-state nonequilibrium expectation~\cite{iop-qpc}, 
\begin{equation} 
\langle 6 \, {\rm operators} \rangle^{\rm sn} := \langle\Psi_0|c_{k'\uparrow}^{R\dagger} (6 \, {\rm operators})  d_\uparrow |\Psi_0 \rangle,
\label{non-eq}
\end{equation}
where the operators \(d_\uparrow\) and \(c_{k'\uparrow}^{R\dagger}\) represent the initial and final events of a cyclic steady-state process, respectively, and 
the ground state \(|\Psi_0 \rangle\) describes a combined electron configuration of the left and right reservoirs at the Fermi level. 
For the reverse flow, $c_{k'\uparrow}^{L\dagger}$ should be used instead of $c_{k'\uparrow}^{R\dagger}$ in equation~(\ref{non-eq}).

These eight selected operators can be rearranged as 
$(d_\uparrow^\dagger c_{k\downarrow}^{L\dagger}) (c_{k\uparrow}^L d_\downarrow) (c_{k'\uparrow}^{R\dagger}d_\downarrow^\dagger) (c_{k\downarrow}^L d_\uparrow)$, 
which describes the following sequence:
(left singlet formation) $\rightarrow$ (singlet cotunneling) $\rightarrow$
(new singlet formation on the left) $\rightarrow$ (spin exchange), as shown in the upper panel of figure~\ref{fig2}.

Meanwhile, the commutators in $\gamma^{LR}_{S}$ and $\gamma^{LR}_{A}$ from equations~(\ref{gamma-S}) and~(\ref{gamma-A}), respectively, are written as
\begin{eqnarray}
&&[j^{-L}_{d\downarrow},j^{+R}_{d\downarrow}]=i\sum_{k\in
L}\sum_{k'\in R}\tilde{V}^L\tilde{V}^R(c^{R\dagger}_{k'\downarrow}d_\downarrow d^\dagger_\downarrow c^L_{k\downarrow} \nonumber \\
&+&c^{L\dagger}_{k\downarrow}d_\downarrow d^\dagger_\downarrow c^R_{k'\downarrow}-d^\dagger_\downarrow c^R_{k'\downarrow}c^{L\dagger}_{k\downarrow}d_\downarrow-d^\dagger_\downarrow c^L_{k\downarrow}c^{R\dagger}_{k'\downarrow}d_\downarrow),
\label{gamma-SS}
\end{eqnarray}
and 
\begin{eqnarray}
&&[j^{-L}_{d\downarrow},j^{-R}_{d\downarrow}]=-\sum_{k\in L}\sum_{k'\in R}\tilde{V}^L\tilde{V}^R(c^{R\dagger}_{k'\downarrow}d_\downarrow d^\dagger_\downarrow c^L_{k\downarrow} \nonumber \\
&-&c^{L\dagger}_{k\downarrow}d_\downarrow d^\dagger_\downarrow c^R_{k'\downarrow}+d^\dagger_\downarrow c^R_{k'\downarrow}c^{L\dagger}_{k\downarrow}d_\downarrow-d^\dagger_\downarrow c^L_{k\downarrow}c^{R\dagger}_{k'\downarrow}d_\downarrow ).
\label{gamma-Anti}
\end{eqnarray}

Depending on the direction of the applied bias, the first two terms on the right-hand side of equations (\ref{gamma-SS}) and (\ref{gamma-Anti}) are used to construct entangled-state tunneling between the left and right reservoirs by combining terms found in the numerators of equation (\ref{gamma-S}) or (\ref{gamma-A})—specifically $c_{k\uparrow}^L d^\dagger_\uparrow$ or $c_{k\uparrow}^R d^\dagger_\uparrow$—with two additional terms arising from the definition of the nonequilibrium expectation, namely $c_{k'\uparrow}^{R\dagger}d_\uparrow$ or $c_{k'\uparrow}^{L\dagger}d_\uparrow$.

entangled-state tunneling between the left and right reservoirs by combining with four $\uparrow$-spin operators: $c_{k\uparrow}^L d^\dagger_\uparrow$ or $c_{k\uparrow}^R d^\dagger_\uparrow$ from the numerator of equation (\ref{gamma-S}) or (\ref{gamma-A}) and another two $c_{k'\uparrow}^{R\dagger}d_\uparrow$ or $c_{k'\uparrow}^{L\dagger}d_\uparrow$ from the definition of nonequilibrium expectation, depending on the direction of applied bias.  

Specifically, for the direction of bias depicted in figure~\ref{fig2}, combining $c^{R\dagger}_{k'\downarrow}d_\downarrow d^\dagger_\downarrow c^L_{k\downarrow}$ in the first term of equations (\ref{gamma-SS}) and (\ref{gamma-Anti}) with $c_{k\uparrow}^L d^\dagger_\uparrow c_{k'\uparrow}^{R\dagger}d_\uparrow$ yields a fourth-order hybridization process of unidirectional entangled-state tunneling by rearranging these eight operators.
The resulting form is 
$(d_\uparrow^\dagger c_{k'\downarrow}^{R\dagger}) (c_{k\uparrow}^L d_\downarrow) (c_{k'\uparrow}^{R\dagger}d_\downarrow^\dagger)$
$(c_{k\downarrow}^L d_\uparrow)$, which represents a sequence: (left singlet formation) $\rightarrow$ (singlet cotunneling) $\rightarrow$
(new singlet formation on the left) $\rightarrow$ (singlet cotunneling), as illustrated in the lower panel of figure~\ref{fig2}.

The second terms in equations~(\ref{gamma-SS}) and (\ref{gamma-Anti}) have the same operator structure as the first terms, except that the reservoir indices $L$ and $R$ are interchanged. 
The two contributions differ in sign. 
This shows that $\gamma^{LR}_{S}$ and $\gamma^{LR}_{A}$ describe the symmetric and antisymmetric combinations, respectively, of leftward and rightward flows between the two reservoirs. 
Consequently, at equilibrium, i.e., at zero bias, $\gamma^{LR}_{A}=0$. 
At nonzero bias, the entangled-state tunneling becomes unidirectional, leading to $\gamma^{LR}_{A}=\gamma^{LR}_{S}$.

\section{Self-energy coefficients $\eta_{pq}$ }
\setcounter{section}{6}
The matrix reduction technique explained in appendix D arrives at a conclusion, i.e., equation (\ref{eq:M-matrix11}):
\begin{equation}	
({\bf M}_{dd}-	{\bf M}_{Ld}{\bf M}_{LL}^{-1}{\bf M}_{dL} -
	{\bf M}_{Rd}{\bf M}_{RR}^{-1}{\bf M}_{dR}){\bf C}^d={\bf I}_5
\label{eq:M-matrix1}
\end{equation}
with $-{\bf M}_{Ld}{\bf M}_{LL}^{-1}{\bf M}_{dL}-{\bf M}_{Rd}{\bf M}_{RR}^{-1}{\bf M}_{dR}=\eta_{pq}[i{\bf \Sigma}_0^L(\omega)+i{\bf \Sigma}_0^R(\omega)]$, where $i{\bf \Sigma}_0^\nu(\omega)=i(\tilde{V}^{\nu})^2\Sigma_{\bf k}[1/(\omega-\epsilon_{\bf k})]$.

Since ${\bf M}=$z{\bf I}$+i{\bf L}$, we have ${\bf M}_{Ld}=i{\bf L}_{Ld}$ and ${\bf M}_{dL}=i{\bf L}_{dL}$, which were addressed in appendix C.3 and shown in figure~\ref{figD1}.
Calculation of $-{\bf M}_{Ld}{\bf M}_{LL}^{-1}{\bf M}_{dL}$ using figure~\ref{figD1} yields $\eta_{pq}[i{\bf \Sigma}_0^L(\omega)+i{\bf \Sigma}_0^R(\omega)]$ with coefficients \(\eta_{pq}\) as follows: 
\begin{eqnarray*}
	\quad \eta_{11}&=&|\xi_{-}^{LL}|^2, \quad \eta_{12}=\xi_{-}^{LL*}\xi_{+}^{LL},
		\quad \eta_{14}=\xi_{-}^{LL*}\xi_{+}^{LR}, \\ 
               \quad \eta_{15}&=&\xi_{-}^{LL*}\xi_{-}^{LR}, 
		 \quad \eta_{22} =|\xi_{+}^{LL}|^2,
		 \quad \eta_{24}=\xi_{+}^{LL*}\xi_{+}^{LR}, \\ 
	\quad \eta_{25}&=&\xi_{+}^{LL*}\xi_{-}^{LR}, \quad
	\eta_{44} =|\xi_{+}^{LR}|^2, \quad \eta_{45}=\xi_{+}^{LR*}\xi_{-}^{LR}, \\
		\quad \eta_{55}&=&|\xi_{-}^{LR}|^2, \quad \eta_{33}=1,
\end{eqnarray*}
where $\eta_{qp}=\eta_{pq}^*$ for $p\neq q$. 
In contrast, $-{\bf M}_{Rd}{\bf M}_{RR}^{-1}{\bf M}_{dR}$ gives the same $\eta_{pq}$ with $L$ and $R$ exchanged.
Half-filling, as is the case in this study, gives rise to real $\xi^{\nu\nu'}_{\mp}$, resulting in real and positive $\eta_{pq}$.

Subsequently, we discuss the standard value, $\eta_{pq}=1/4$, used for the phenomenological determination of $\eta_{pq}$ in the main text. 
The differential conductance formula, equation~(\ref{Conductance}), yields Coulomb peaks at 
$\pm U/2$ with conditions $U_{j^\pm}^{L,R}=U/4$ and $\gamma^{L,R}=\gamma^{LR}_{S,A}=0$ at the atomic limit, namely ${\rm Im}\Sigma(\omega) \rightarrow 0$. 

Comparing the expressions of \(U_{j^\mp}^\nu\) in equation~(\ref{eq:Ualphaj}) and $\xi_{\mp 1}$ in equation~(\ref{eq:xi-1}) with $\langle(\delta n_{d\downarrow})^2\rangle^{1/2}=1/2$ for half-filling, i.e.,
$$ \xi^{\nu\nu'}_{\mp 1}=\frac{\langle i(1-2n_{d\uparrow})
	[n_{d\downarrow},j^{\mp \nu'}_{d\downarrow}]\rangle+i(1-2\langle n_{d\downarrow}\rangle)\langle
	j^{\mp\nu'}_{d\downarrow}\rangle}{\langle(\delta j^{\mp \nu'}_{d\downarrow})^2\rangle^{1/2}} $$
gives rise to 
$$\xi^{\nu\nu'}_{\mp 1}=2U_{j^\mp}^{\nu'}/U,$$
which yields $\xi^{\nu\nu'}_{\mp 1}=1/2$, namely the part without left--right entanglement, for $U_{j^\mp}^{\nu'}=U/4$.

As shown above, since $\eta_{pq}$ is given by the product of two $\xi^{\nu\nu'}_{\mp}$'s, we obtain the standard value of $\eta_{pq}=1/4$ in the absence of entanglement effects. 
 
Meanwhile, as discussed in appendix C.3, the contribution from left--right entanglement is denoted by $\xi^{\nu\nu'}_{\mp 2}$. 
These satisfy the inequality relation given in equation~(\ref{eq:inequality}), which naturally leads to the following inequality for $\eta_{pq}$:
$\eta_{11}=\eta_{15}=\eta_{55}<\eta_{12}=\eta_{14}=\eta_{25}=\eta_{45}<\eta_{22}=\eta_{24}=\eta_{44}$.
In the main text, we determined the values of $\eta_{pq}$ phenomenologically using this relation and its standard value $1/4$.

\end{document}